\documentclass{IEEEjmw}

\usepackage[colorlinks,urlcolor=blue,linkcolor=blue,citecolor=blue]{hyperref}

\usepackage{color,array}

\usepackage{graphicx}
\usepackage{booktabs}
\usepackage{amsmath,amssymb}
\usepackage{fixltx2e}
\usepackage{float}
\usepackage{comment}
\usepackage{upgreek}
\usepackage{dblfloatfix}
\usepackage{float}
\usepackage{gensymb}
\usepackage{algorithm,algorithmic}
\usepackage{cite}

\newcommand{\bldW}{\mathbf W}
\newcommand{\bldw}{\mathbf w}
\newcommand{\bldDlt}{\boldsymbol \Delta}
\newcommand{\blddlt}{\boldsymbol \delta}
\newcommand{\bldzero}{\boldsymbol 0}

\jvol{00}
\jnum{XX}
\paper{1234567}
\pubyear{20XX}
\receiveddate{Month~XX, 20XX}
\accepteddate{Month~XX, 20XX}
\publisheddate{Month~XX, 20XX}
\currentdate{Month~XX, 20XX}
\doiinfo{JMW.20XX.Doi Number}

\begin{document}

\sptitle{Article Category}

\title{Distributed Beamforming Using Decentralized Time Synchronization in a Six-Element Array} 

\editor{The associate editor coordinating the review of this manuscript and approving it for publication\break was F. A. Author.}

\author{Naim Shandi\affilmark{1} (Graduate Student Member, IEEE)}

\author{Jason M. Merlo\affilmark{1}  (Graduate Student Member, IEEE)}

\author{Jeffrey A. Nanzer\affilmark{1} (Senior Member, IEEE)}

\affil{Department of Electrical and Computer Engineering, Michigan State University, East Lansing, MI 48824 USA}

\corresp{CORRESPONDING AUTHOR: Jeffrey A. Nanzer (e-mail: \href{mailto:Nanzer@msu.edu}{Nanzer@msu.edu}).}
\authornote{This work was supported in part by Office of Naval Research under Grant \#N00014-20-1-2389, the National Science Foundation under Grant \#1751655, and Google under the Google Research Scholar program.}

\markboth{PREPARATION OF PAPERS FOR IEEE JOURNAL OF MICROWAVES}{Naim Shandi {\itshape ET AL}}

\begin{abstract}
We demonstrate distributed beamforming and beamsteering from a six-node distributed phased array using fully wireless coordination with decentralized time synchronization. In wireless applications such as distributed beamforming, high-accuracy time synchronization across the array is crucial for high coherent gain. The decentralized time synchronization method employed is based on the average consensus algorithm and the two-way time transfer method presented in our previous work, which achieved picosecond-level time synchronization with a cabled frequency reference. The system presented in this paper utilizes a centralized wireless frequency transfer method to achieve wireless frequency syntonization in a fully wireless coordination and a distributed computing system architecture. We experimentally evaluate system performance through beamforming and beamsteering to a receiver 16.3 m away from the six-node non-uniformly distributed antenna array, achieving an average coherent gain of 98\% of the ideal gain at a carrier frequency of 1.05 GHz. The average time synchronization accuracy achieved was less than 36 ps.
\end{abstract}


\maketitle

\section{INTRODUCTION}
\label{intro}

\begin{figure}
\centerline{\includegraphics[width=18.5pc]{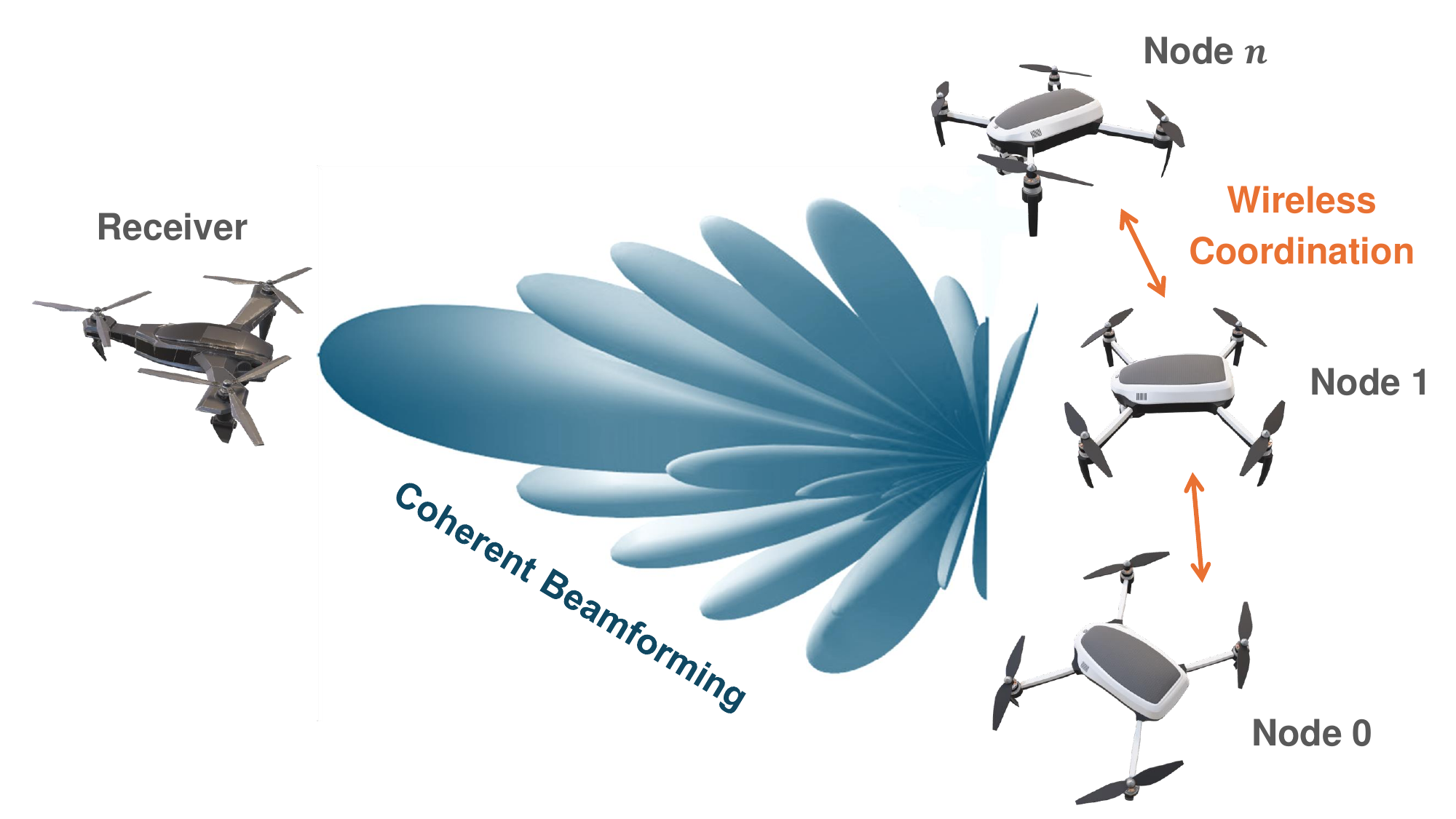}}
\caption{Example of coherent beamforming in a distributed antenna array system mounted on drones.}
\label{DAA}
\end{figure}

Coherent distributed antenna arrays are increasingly important in communication systems and remote sensing applications because of their ability to dramatically enhance signal gain, improve energy efficiency, and increase signal coverage. New capabilities can be enabled, such as distributed collaborative beamforming (Fig.\ \ref{DAA}), which increases the signal-to-noise ratio (SNR), enhances security, and reduces interference by directing signals to the receiver while reducing the power scattered in other directions \cite{4785387,nanzer2021distributed}.  Unlike traditional monolithic antenna arrays, the architecture of coherent distributed arrays (CDAs) enables improved signal power and quality in challenging environments, with a scaling factor of $N^2M$ for an array with $N$ transmitters and $M$ receivers (each node equipped with its own power supply). These advantages for DAAs, in addition to their resilience to single-node failure and low maintenance cost, have increased their applicability to emerging technologies such as Internet-of-Things (IoT) devices \cite{mi13091481}, infrastructure for 6G communication networks \cite{10475582,9798882}, and distributed remote sensing networks \cite{9376348}. 

The distributed nature of CDAs introduces challenges in coordinating the system's spatially separated elements to a sufficiently accurate level to achieve and maintain coherent distributed operation. Time, phase, and frequency synchronization at a fraction of the radio frequency carrier wavelength are essential to enable a highly coherent summation of the transmitted signals at the target location~\cite{nanzer2021distributed,8378649}
. The challenges of CDA coordination have motivated the development of new methods and the improvement of existing methods that can achieve the precise coordination required for high gain distributed beamforming by continuously compensating for clock drift, time offsets, latency, and varying delays. The methods are mainly classified into two groups of systems. Closed-loop systems depend on acquiring information and feedback from the receiver to achieve coordination and collaborative beamforming \cite{4202181,4542555,4641607,bidgare2012implementation}. Open-loop systems require no feedback from the receiver~\cite{abari2015airshare,prager2020wireless,alemdar2021rfclock}
; these systems are required in many applications such as remote sensing and radar where feedback from the destination is unavailable.

CDA coordination methods can also be classified, based on the type of network topology, into centralized and decentralized topologies. In the centralized topology, the array depends on one primary or reference node to provide information to the other nodes in the array, which provides a fairly simple architecture but makes the topology prone to single-point failure. Some recent research on these methods is presented in~\cite{merlo2022wireless,prager2020wireless}. Recently, interest has shifted to the development of decentralized methods capable of providing high precision CDA coordination and addressing the drawbacks of the centralized approach. The decentralized methods allow for information sharing between the neighboring elements of a CDA without depending on a reference node, thus eliminating the possibility of single-point failure and improving network robustness~\cite{wang,ouassal2021decentralized,140489,10600440,10443654}.
Developing a CDA system capable of achieving and maintaining high accuracy coordination in a fully wireless communication link is a challenging problem. Previously proposed systems generally depended on an external reference such as the global positioning system (GPS) to lock their internal oscillators and achieve subnanosecond centralized synchronization accuracy (e.g.,~\cite{prager2020wireless}). Recent publications focused on developing digital synchronization techniques to eliminate the need for an external clock reference (see \cite{10600440,10443654}). In \cite{10600440,10443654} the authors demonstrated coordination without the need for an external frequency or time reference. In \cite{merlo2024past} we demonstrated beamforming in a cluttered environment using a centralized topology for high accuracy digital time synchronization, also achieving beamforming without the need for external time or frequency references with a small array.

In this work we present distributed beamforming and beamsteering in a fully wireless coordinated distributed phased array system. The time synchronization method was decentralized using the average consensus algorithm at a fraction of the RF carrier wavelength without the need for external time or frequency references. Frequency syntonization is centralized but fully wireless; future work will decentralize the syntonization approach as well. The present paper significantly expands on our previous work in~\cite{merlo2024past} and~\cite{shandi2024} by: (1) scaling the system from three elements to six elements, experimentally demonstrating fully wireless beamforming and beamsteering in a large network of distributed antenna arrays; (2) validating the decentralized time synchronization method through performing collaborative beamforming and beamsteering in a cluttered environment, achieving a mean coherent gain of 98\% with an average time synchronization accuracy of less than 36 ps.  
The rest of the paper is organized as follows. Section \ref{Decentralized_time_sync} gives a detailed discussion of the decentralized time synchronization method, followed by a study of the effects of time alignment errors on the distributed coherent beamforming gain. Discussion of the frequency syntonization technique employed in this work is also provided in Section \ref{freq_syn}. Section \ref{beamforming} describes distributed beamforming and beamsteering procedures for the nonuniform distributed linear array, highlighting the array's main characteristics and describing the calibration process. The experimental evaluation is detailed in Section \ref{experimental_eval}, including a discussion of the results, followed by the conclusion in Section \ref{conclustions}.

\section{DECENTRALIZED TIME SYNCHRONIZATION}
\label{Decentralized_time_sync}
\subsection{SYSTEM MODEL}
\label{sys_model}

Achieving high coherent gain for distributed antenna array beamforming requires precise time, phase, and frequency alignment of the electrical signals at all the antenna elements (which may be transmitters, receivers, or both) in the array. Hence it is crucial to establish a mathematical model that closely approximates the time of each clock in the system. The time model implemented in this work is based on that developed in~\cite{merlo2022wireless}, where the clock drift at node $n$ over a short time interval is approximately linear and therefore the time at the same node can be represented by a linear model. 

To achieve time synchronization across the array, the relative time offset estimated at node $n$ can be modeled as a function of the global time $t$ as
\begin{equation}\label{time_model}
T^{(n)}(t) = \alpha^{(n)}(t) t + \beta^{(n)}+\delta^{(n)}(t) + \nu^{(n)}(t),
\end{equation}
where $\alpha(t)$ is the relative clock frequency, $\beta$ is the static clock time offset or the initial time offset at the RF front-end, $\delta(t)$ is the dynamic clock time bias or the time-varying bias which accounts for factors such as the temperature-dependent delays through the RF components of the system, and $\nu(t)$ is a zero-mean noise function that includes the effects of noise sources contributing, e.g., thermal noise or flicker noise \cite{pozar2005microwave}. The aim is to continuously estimate and correct for the time biases $\beta$ and $\delta(t)$ at the RF front-end in order to achieve the time alignment accuracy needed for high coherent gain. Note that the LO frequency drift is corrected using the frequency transfer method discussed in Section \ref{freq_syn}. Therefore, in the time model $\alpha(t) \equiv 1$.

The LO phase at node $n$ can be written in terms of the time model in \eqref{time_model} as
\begin{equation}\label{phase}
    \Phi^{(n)}(t) = e^{j2\pi f_c T^{(n)}(t)+ \phi^{(n)}_0}
\end{equation}
where $f_c$ is the carrier frequency and $\phi^{(n)}_0$ is the static phase which can be estimated and compensated for (Section \ref{beamforming}-\ref{calibration}).

\subsection{INTER-NODE CLOCK OFFSET AND RANGE ESTIMATION}
\label{offset_range_est}

\begin{figure}[t!]
\centering
\includegraphics[width=1\columnwidth]{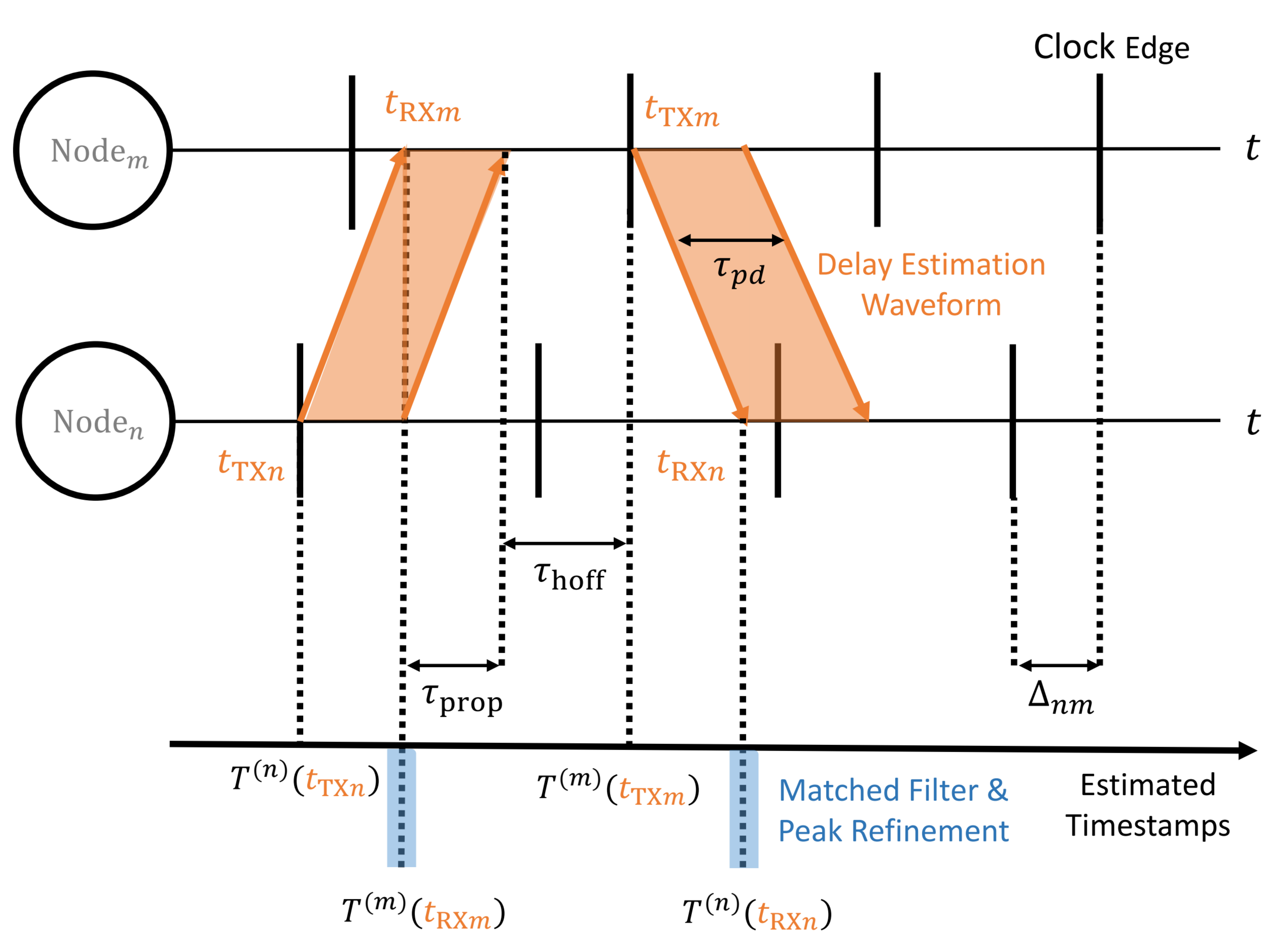}
\caption{TWTT method applied between nodes $n$ and $m$ in an unsynchronized array. The time offset $\Delta_{nm}$ is shown as the difference between the clock edges. The delay estimation waveform is sent from node $n$ to node $m$, and the times of transmission and reception are estimated and saved. Node $m$ sends back a delay estimation waveform to node $n$. After all nodes send and receive, the timestamps are exchanged between the nodes through the network for time offset and range estimation. Here $\tau_{\text{prop}}$ is the propagation delay, which is the time between the transmission by one node and the reception of the same signal by the other node; $\tau_{\text{hoff}}$ describes the time between receipt of the signal by node $m$ and transmission of the signal back to node $n$; and $\tau_{pd}$ is the pulse duration.} 
\label{TWTT}
\end{figure}

The two-way time transfer (TWTT) method is used in this work to estimate the time offset and the range between each pair of nodes in the array. In this method a delay estimation waveform is transmitted between nodes $n$ and $m$ as described in Fig.\ \ref{TWTT} and four timestamps---$T^{(n)}(t_{\text{TX}n})$ and $T^{(m)}(t_{\text{TX}m})$ for transmission, and $T^{(n)}(t_{\text{RX}n})$ and $T^{(m)}(t_{\text{RX}m})$ for reception---are saved and exchanged between the nodes. Here, $t_{\text{TX}n}$ and $t_{\text{RX}n}$ are the time of transmission and reception, respectively. The accuracy of the time alignment process depends mainly on how accurately the times of arrival $T^{(n)}(t_{\text{RX}n})$ and $T^{(m)}(t_{\text{RX}m})$ can be estimated. Estimating the timestamps of the transmission can be done by scheduling the transmission to start at the digital-to-analog converter clock edge, resulting in estimation accuracy limited primarily by the sampling clock jitter. However, it is uncertain when the signal is received; this can be at any instant between the analog-to-digital converter (ADC) clock edges, which makes estimating the time of arrival (ToA) of the signal more challenging. For this purpose, the two-stage peak refinement process was used to estimate the ToA of the received signal to within a single clock cycle of the ADC by applying a matched filter and then refining the peak estimate through quadratic least-squares interpolation and a residual bias correction table as described in \cite{merlo2022wireless}. 
In this work the peak estimation and refinement process for each node happens at each host computer. The saved timestamps are transmitted using TCP/IP network packets over a Wi-Fi link to the time synchronization controller host (running on one of the nodes in the array, which is chosen arbitrarily).

The inter-node clock offset can be estimated to sub-sample level in every synchronization period (equivalently, in each iteration $k=1,2,\ldots$ of the decentralized algorithm) from the four saved timestamps using~\cite{merlo2022wireless}
\begin{equation}
    \Delta_{nm}(k)=\frac{1}{2}\left[T^{(n\to m)}(k)-T^{(m\to n)}(k)\right],
\end{equation}
where 
$$
 T^{(n\to m)}(k)=T^{(m)}[t_{\text{RX}m}(k)]-T^{(n)}[t_{\text{TX}n}(k)]
$$
and
$$
 T^{(m\to n)}(k)=T^{(n)}[t_{\text{RX}n}(k)]-T^{(m)}[t_{\text{TX}m}(k)]. 
$$
Additionally, the inter-node range at the $k$th iteration can be estimated to sub-centimeter precision by averaging the same set of times then multiplying the result by the speed of light 
\begin{equation}
    R_{nm}(k)=\frac{c}{2}\left[T^{(n\to m)}(k)+T^{(m\to n)}(k)\right].
\end{equation}
An important assumption has been made regarding the nature of the communication channel: in order to cancel the channel's impact on the TWTT method, the channel must be reciprocal during the synchronization period. This assumption is satisfied for a relatively static network~\cite{levine2008review}.

\subsection{Average Consensus Algorithm}
\label{ave_consensus}

Two types of network topologies, centralized and decentralized, can be implemented to continuously compensate for the time biases estimated with the TWTT method in order to synchronize the internal clocks of all nodes in the network and maintain a high level of coordination accuracy suitable for beamforming applications. As discussed in Section \ref{intro}, the decentralized network topology is generally preferable for its robust performance compared to that of the centralized topology. One algorithm used to achieve decentralized networking is the average consensus algorithm, where every node in the array estimates the time offsets with its neighboring nodes and computes and compensates for the average bias relative to the nodes linked to it. After multiple iterations, all nodes in the array will achieve time synchronization with clock timing equal to the average of the initial clock states of the nodes in the network. To illustrate the average consensus algorithm, we consider an $N$-node distributed network represented by an undirected graph model $G = \{\mathcal{N}, \mathcal{E}\}$ where $\mathcal{N}=\{1, 2, 3, \ldots, N\}$ and $\mathcal{E}$ is a set of edges with bidirectional communication occurring along each edge. The notation $(n,m) \in \mathcal{E}$ means that nodes $n$ and $m$ are connected~\cite{ouassal2021decentralized}. The implementation of the average consensus suggests that the average time offset at node $n$ is $\bldw_n \cdot \blddlt_n$ where $\bldw_n$ is the $n$th row vector of a weighting matrix $\bldW$ (described further below) and $\blddlt_n$ is the $n$th column vector of $\bldDlt$, the matrix of time biases. 

The implementation of the average consensus algorithm is based on the weighting or mixing matrix $\bldW=~[ \bldw_1 \bldw_2 \cdots \bldw_n]^T$, which is an $n \times n$ real Metropolis–-Hastings constant edge matrix. The $n$th row of $\bldW$ represents the local connectivity information (edges of the graph) for node $n$ with itself and its neighboring nodes, with zero entries for disconnected channels and unit entries for the connected channels including the self loops. The advantages of the average consensus algorithm are: (1) it is dependent on local information only (no global information sharing is needed); (2) it ensures convergence to the average of the initial states when the mixing matrix is symmetric ($\bldW = \bldW^T$), doubly stochastic (the elements of each row and column sum to unity), and decentralized ($\bldw_{nm} = 0$ when nodes $n$ and $m$ are disconnected and $n \neq m$)~\cite{boyd}. 

In the present work, decentralized time synchronization is considered for the case of a fully connected network with a fixed network graph (i.e., the inter-node connectivity is static during the experiment). In this case the local time at node $n$ is updated during every iteration after estimating the time offsets between node $n$ and all the nodes in the system. The average bias at iteration $k$ is computed and compensated for by
\begin{equation}
    T^{(n)}(k+1)= T^{(n)}(k) + \bldw_n \cdot \blddlt_n(k).
\end{equation}
Note that $\bldw_n$ does not depend on $k$ for static network connectivity, while $\blddlt_n$ is estimated in every iteration using the TWTT method. For a fully connected network, the time at each node converges (i.e., reaches a steady state where the inter-node time offsets are small, $\bldDlt \approx \bldzero$) to the average initial time of all the nodes in the array after two iterations~\cite{shandi2024}. 

%
%
%
%
%
%

\subsection{Effect of Timing Error on Coherent Beamforming in Radar Systems}
\label{time_error}

The effect of time synchronization errors on the coherent operation of distributed systems is detailed in \cite{8378649}. The authors modeled the relative matched filter gain as a function of timing error, expressed as a percentage of the pulse duration of different waveforms including the continuous-wave (CW) pulse train used as the beamforming waveform in the present work. The coherent matched filter gain of the system was evaluated as a function of the timing errors. The signal model that considered only the timing error for the received signal (transmitted signal reflected by a target) was represented as 
\begin{equation}\label{time_error}
    r_{\text{RX}}(t)=\sum_{n=1}^Nu(t-\delta_n) \, e^{j2\pi f_ct},
\end{equation}
where $\delta_n$ is the time synchronization error and $u$ is a unit-amplitude rectangular pulse. A matched filter was applied to the received signal and the ideal transmitted signal 
\begin{equation}\label{ideal}
    r_{\text{TX}}(t)=u(t) \, e^{j2\pi f_ct}, 
\end{equation}
and the peak of the matched filter was then computed in a Monte Carlo simulation as a function of the standard deviation of the timing error normalized to the CW pulse width $T$. In the case of a CW pulse train \eqref{time_error} and \eqref{ideal} become
\begin{align*}
    S_{\text{TX}}(t)&=\Pi \left( \frac{t}{T} \right) e^{j2\pi f_ct},\\[4pt]
    S_{\text{RX}}(t)&=\Pi \left( \frac{t-\delta_n}{T} \right) e^{j2\pi f_ct},
\end{align*}
where
\begin{align}
    \Pi \left( \frac{t}{T} \right)=\begin{cases}
        1, \quad &  |t|< \frac{T}{2},\\
        0, \quad & \text{otherwise}.
    \end{cases}
\end{align}
The statistical evaluation of the timing error was quantified through computing the probability that the matched filter coherent gain $G_c$ is at least 90\% of the ideal matched filter gain value (unity gain), that is, $P(G_c \geq 0.9)$. 

\begin{figure}[t!]
\centering
\includegraphics[width=1\columnwidth]{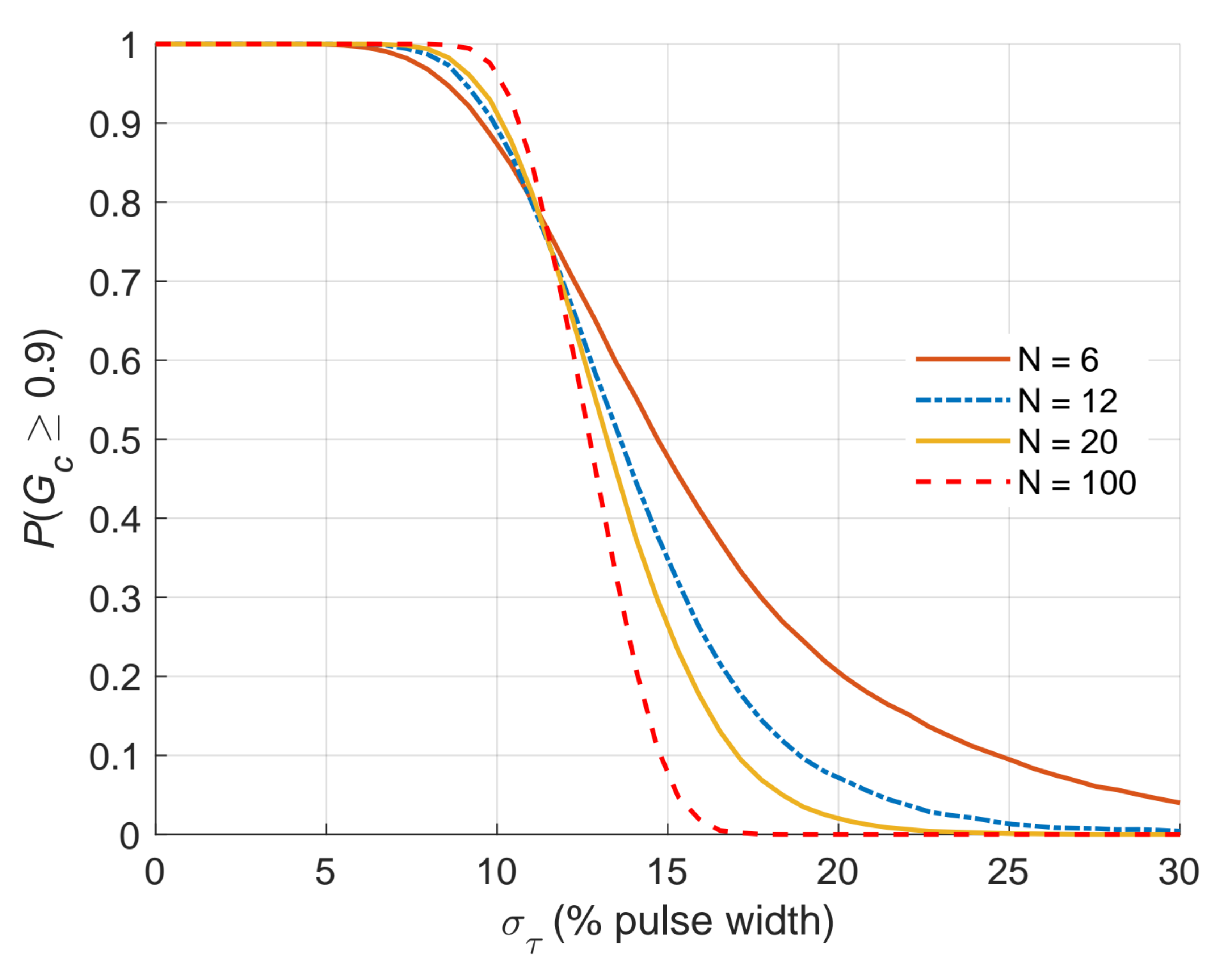}
\caption{Timing error evaluation for antenna array sizes of 6, 12, 20, and 100 nodes. Through 1000 Monte Carlo simulations, the relative coherent gain was computed as a function of the timing error for the CW pulse and compared for different array sizes. For a six-node array, to achieve at least 90\% of the ideal gain with probability of 0.9, the standard deviation of the timing error must be under 9.5\% of the CW pulse width.}
\label{ETE}
\end{figure}

In Fig.\ \ref{ETE}, this probability is computed through 1000 Monte Carlo simulations for different array sizes and plotted as a function of the standard deviation of the timing error as a percentage of the CW pulse width in the range 0--30\%. Note that for an array size of at least 20 nodes, the probability of achieving greater than 90\% coherent gain above 90\% of the time was achieved when the standard deviation of the tolerable timing error was $\sim$10\% of the pulse width, meaning the requirement is less stringent for larger arrays. As the array size decreased to six nodes, the effect of the timing error becomes more significant. For a six-node array, in order to achieve a probability of at least 90\% of achieving coherent gain above 90\%, the standard deviation of the timing error must be less than 9.5\% of the CW pulse width. Also note that when the standard deviation of the timing error increased beyond 10\% of the pulse width, the probability of achieving 90\% coherent gain decreased (the decrease was slower for smaller array sizes). 

The results of this analysis highlight the importance of the time alignment accuracy in distributed beamforming by showing the significance of the effect of timing error on the probability of achieving high coherent gain in distributed phased arrays. In addition to the hardware limitations on phase alignment accuracy, which will be discussed in following sections, the simulation result suggests maintaining a range of timing accuracy standard deviation of 0--9.5\% of the CW pulse width to achieve a probability of at least 90\% that the coherent gain will exceed 90\% during beamforming.   

%



\section{WIRELESS FREQUENCY SYNTONIZATION}
\label{freq_syn}
In addition to accurate time and phase alignment, precise frequency syntonization---where all the local oscillators (LOs) in the network operate at the same carrier frequency---is required to achieve high coherent gain for distributed antenna arrays. With the slow drift of LO frequency over time, continuous frequency syntonization is needed to estimate and compensate for the initial frequency offset as well as the time varying stochastic drift. Different methods discussed in the literature enable wireless frequency syntonization of LOs within a network of multiple agents, each having its advantages and limitations. Some of these methods include frequency syntonization using orthogonal frequency-division multiple access using data packets \cite{8649676}, digital frequency syntonization \cite{10443654}, frequency alignment using coupled oscillators \cite{8058715}, optically locked voltage-controlled oscillators \cite{6848566}, global positioning system signals \cite{4319216,prager2020wireless}, and two-tone frequency locking \cite{mghabghab2021open-loop,alemdar2021rfclock}. Generally, frequency and time alignments are accomplished jointly. However, if the time alignment method is accurate enough and is achieved in a relatively short period of time compared to the frequency drift, this drift is minimized, and synchronization is accomplished by adjusting the LO phase only \cite{merlo2022wireless}. 

\begin{figure}[t!]
\centering
\includegraphics[width=1\columnwidth]{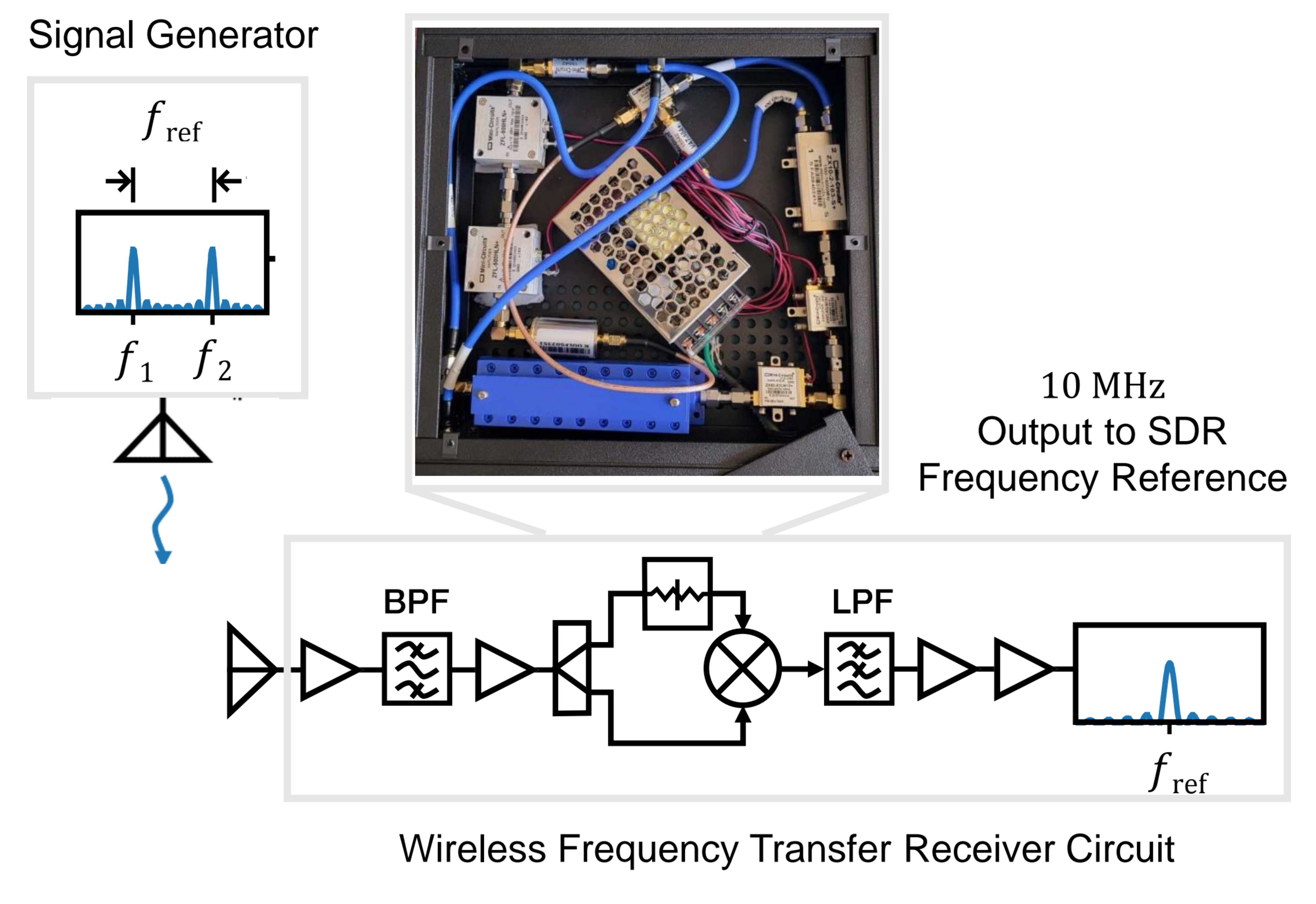}
\caption{Block diagram of the two-tone self-mixing frequency locking receiver. The 10 MHz frequency reference is obtained by restructuring the received two-tone waveform (with 10 MHz tone separation), specifically by splitting it and mixing it with a copy of itself to create a square wave with the desired frequency. Photo of implemented hardware appears at top of figure.}
\label{wft}
\end{figure}

In \cite{shandi2024} we evaluated the accuracy of decentralized time synchronization in a lab environment using the conventional method of frequency locking the LOs of nodes to the PLL of a primary node, providing a 10 MHz reference signal through cables. In this work we employ the one-way two-tone frequency transfer method to achieve fully wireless coordination with the frequency syntonization needed for aligning the received waveforms within the matched filtering window, thus enabling accurate time delay and range estimation. For this purpose, the self-mixing frequency locking receiver detailed in~\cite{mghabghab2021open-loop,abari2015airshare} is employed, shown in Fig.\ \ref{wft}. To achieve syntonization using this method, the continuous two-tone waveform with 10 MHz bandwidth transmitted by one node in the system is received by the other nodes and processed (split and mixed with itself) at the self-mixing circuit; this produces a 10 MHz square wave required to frequency lock the LOs of the other nodes. Future efforts will apply previously developed decentralized syntonization approaches (e.g.,~\cite{ouassal2021decentralized}) combined with decentralized time synchronization.

\section{DISTRIBUTED BEAMFORMING}
\label{beamforming}
\begin{figure}[t!]
\centering
\includegraphics[width=1\columnwidth]{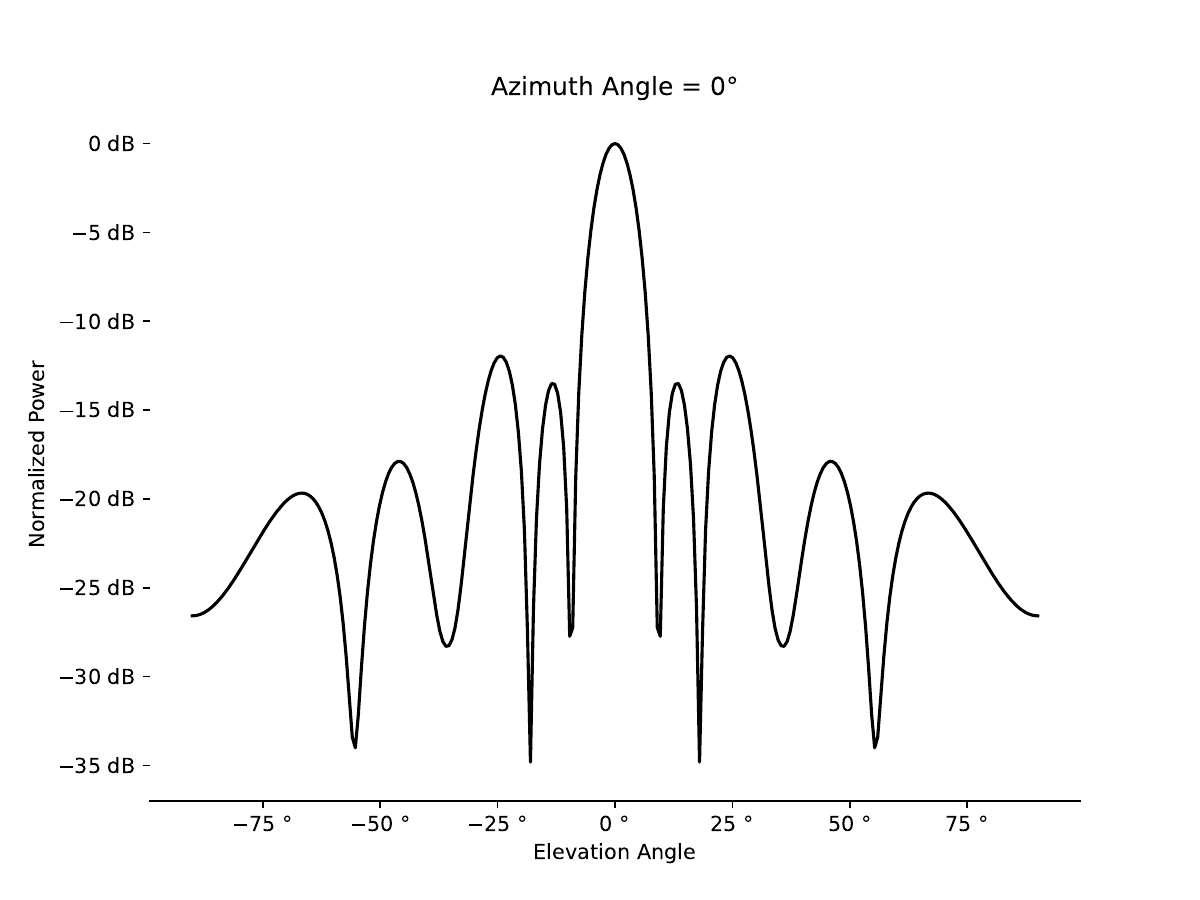}
\caption{Elevation cut of the six-node array power pattern.}
\label{directivity}
\end{figure}

The coordination methods of Section \ref{Decentralized_time_sync} are required to align the electrical states of the nodes in the system to enable distributed beamforming with coherent gain of at least 90\% of ideal (i.e., less than 0.5 dB degradation). Viability of the coordination methods --- specifically the decentralized time synchronization method --- and the scalability of the system performance were assessed via beamforming to a receiver located at broadside, and digital beamsteering from broadside to 45$\degree$. Prior to the experimental evaluation, the array directivity and power pattern were computed to assess the array pattern and thus control the element phase weighting to focus the beam in the desired direction and understand the effect of interference and multipath on the beamsteering results. To simplify the calibration process, the six nodes were positioned along the $x$-axis with node 0 at the origin: the locations were $x = -0.648$, $x = -0.368$, $x=0$, $x = 0.213$, $x =0.551$, and $x =0.813$ meters.
The array element pattern was modeled as a sinc function to closely match the pattern of the 11 dBi L-Com HG72710LP-NF log-periodic antenna used in the beamforming; the horizontal and vertical beamwidths of this antenna are 78$\degree$ and 56$\degree$, respectively. 
Figure \ref{directivity} shows a polar coordinate elevation cut of the 2D normalized power pattern of the array.

\subsection{SYSTEM CALIBRATION}
\label{calibration}
\begin{figure*}[t!]
\centering
\includegraphics[width=2\columnwidth]{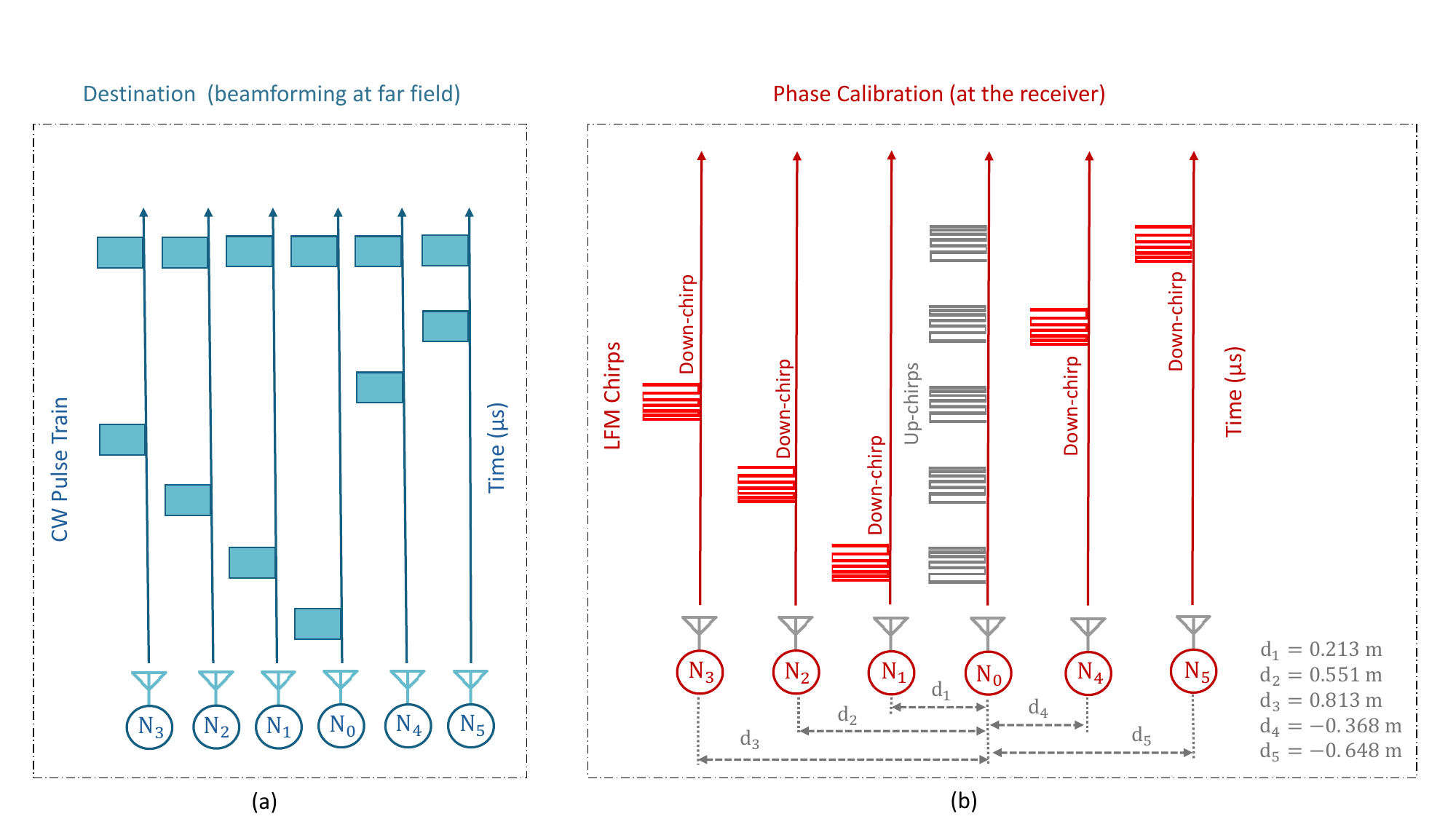}
\caption{Evaluation and calibration processes. (a) The pulse train sequence transmitted by each node, and the node's relative location. All the nodes sent pulse trains simultaneously; however, the amplitudes of each sequence shifted in such a way as to allow the receiving node to capture seven pulses. The first six pulses come from the individual nodes, while the last is the combination of all six pulses. (b) The calibration process and the measured inter-node distances $d_{0n}$ with respect to node 0.}
\label{cw}
\end{figure*}

In order to beamform and steer the beam digitally, system calibration is needed to compensate for the static system delay $\tau_{\text{bf,cal}}$ (i.e., the delay introduced by the signal processing and RF hardware properties) and the phase error $\phi_0$ in \eqref{phase}. In many applications it is preferable to have open-loop beamforming where no feedback from the destination is needed, and these static phases can be calibrated in various ways such as transceiver characterization or hardware feedback loops. In this work, system carrier synchronization is accomplished using an open-loop approach (Section \ref{Decentralized_time_sync}) where no information from the destination is needed. However, for simplicity the beamforming delay and phase calibration are based measurements taken in the far-field at the evaluation receiver, after which no signals from the evaluation receiver are used. Calibration of $\phi_0$ and $\tau_{\text{bf,cal}}$ for the beamforming pulses was accomplished using orthogonal LFM chirps transmitted from the pair of nodes $N_{0n}$ ($n=1,2,\ldots,5$) to the receiver (node 0 was chosen arbitrarily; any node in the array can function as the reference node in this calibration process). Node 0 transmitted a 160~MHz LFM up-chirp while the other nodes transmitted 160~MHz LFM down-chirps in the time scheduled TDM process of Fig.\ \ref{cw}-b. The chirps were received at the destination (a four-channel oscilloscope (Keysight DSOS0804A) with 20~GSa/s sampling rate) and matched filtered twice, with the ideal up-chirp and the ideal down-chirp, to estimate the inter-arrival time delay and phase.
These estimated parameters were added to the transmitted beamforming pulses during the beamsteering experiment after the QLS and peak refinement stage.

In addition to the delay calibration, which must be done once before the evaluation experiment starts, an additional calibration is needed to correct for the static range estimation bias caused by the hardware group delay, signal processing delay, and RF transmission lines~\cite{merlo2024past}. No feedback from the destination is needed in the inter-node range estimation. Instead, the array inter-node distances $d_{0n}$, which are measured with respect to a reference node (in this case node 0), are divided by the speed of light $c$ to form the true inter-node delays. These values are then subtracted from the estimated times of flight $\hat \delta_{0n}$ obtained from the application of TWTT while keeping the nodes at static locations. The resulting values, which are the inter-node time biases 
\begin{equation}
    \delta_{0n,\text{cal}}=\hat \delta_{0n}-d_{0n}/c
\end{equation}
are saved in the calibration file. This calibration compensates for the static delays of the signal processing and those caused by the cables and interconnectors; it needs to be done just once
before the beamsteering experiment \cite{merlo2024past}. 

\subsection{Beamforming and Beamsteering}
\label{beamforming_beamsteering}
Beamforming at angles from broadside ($0\degree$) to $45\degree$ were performed to evaluate the system performance. Two parameters were used for evaluation: (1) the time of arrival estimated at the receiver during beamsteering, and (2) the system gain during beamforming and beamsteering. An LFM waveform was used in the time-of-arrival estimation, while the system gain was estimated using a CW pulse train to evaluate the phase accuracy. The transmitted signal compensates for the carrier phase offsets in \eqref{phase} and imparts beamforming delays, and can be given by
\begin{equation}
   S^{(n)}_{\text{TX}}(t, \theta) = s_a\left(T^{(n)}(t) - \tau^{(n)}_{\text{bf}}(\theta)\right)e^{-j\left[2 \pi f_c \tau^{(n)}_{\text{bf}}(\theta)+\phi^{(n)}_0\right]},
\end{equation}
where $s_a(t)$ is the signal amplitude, which may be complex, and
\begin{equation}
    \tau^{(n)}_{\text{bf}} (\theta) = \left(\frac{r_{0n}}{c} -\delta_{0n,\text{cal}}\right)\sin \theta + \tau^{(n)}_{\text{bf},\text{cal}},
\end{equation}
where $r_{0n}=R_{0n}(k)$ and $\theta$ is the beamsteering angle. 
 
In the evaluation process, the array nodes were scheduled to send LFM waveforms and CW pulse trains in an alternating pattern. Each node was scheduled to send LFM waveforms followed by a train of simple pulses to the receiver, which processed them sequentially. The waveform transmission was scheduled using time division multiple access (TDMA) as shown in Fig.\ \ref{cw}-a. The pulse train was coded such that each node transmitted two pulses, the first at a time slot specific to each node, and the second simultaneously among all nodes, allowing the receiving node to receive six pulses representing the signal power transmitted by each of the nodes, and a seventh pulse representing the combined power resulting from the sum of all the signals scaled by the array pattern. In the beamsteering experiment, the measurement was repeated twice at each angle while steering the beam from $0\degree$ to $45\degree$. The received combined signal power was measured and compared to the ideal signal power (assuming all the nodes were perfectly synchronized in time, phase, and frequency) to estimate the coherent gain of the array, or the ratio of the two powers.

\section{EXPERIMENTAL EVALUATION}
\label{experimental_eval}
\subsection{EXPERIMENTAL CONFIGURATION}

\begin{figure*}[t!]
\centering
\includegraphics[width=1.85\columnwidth]{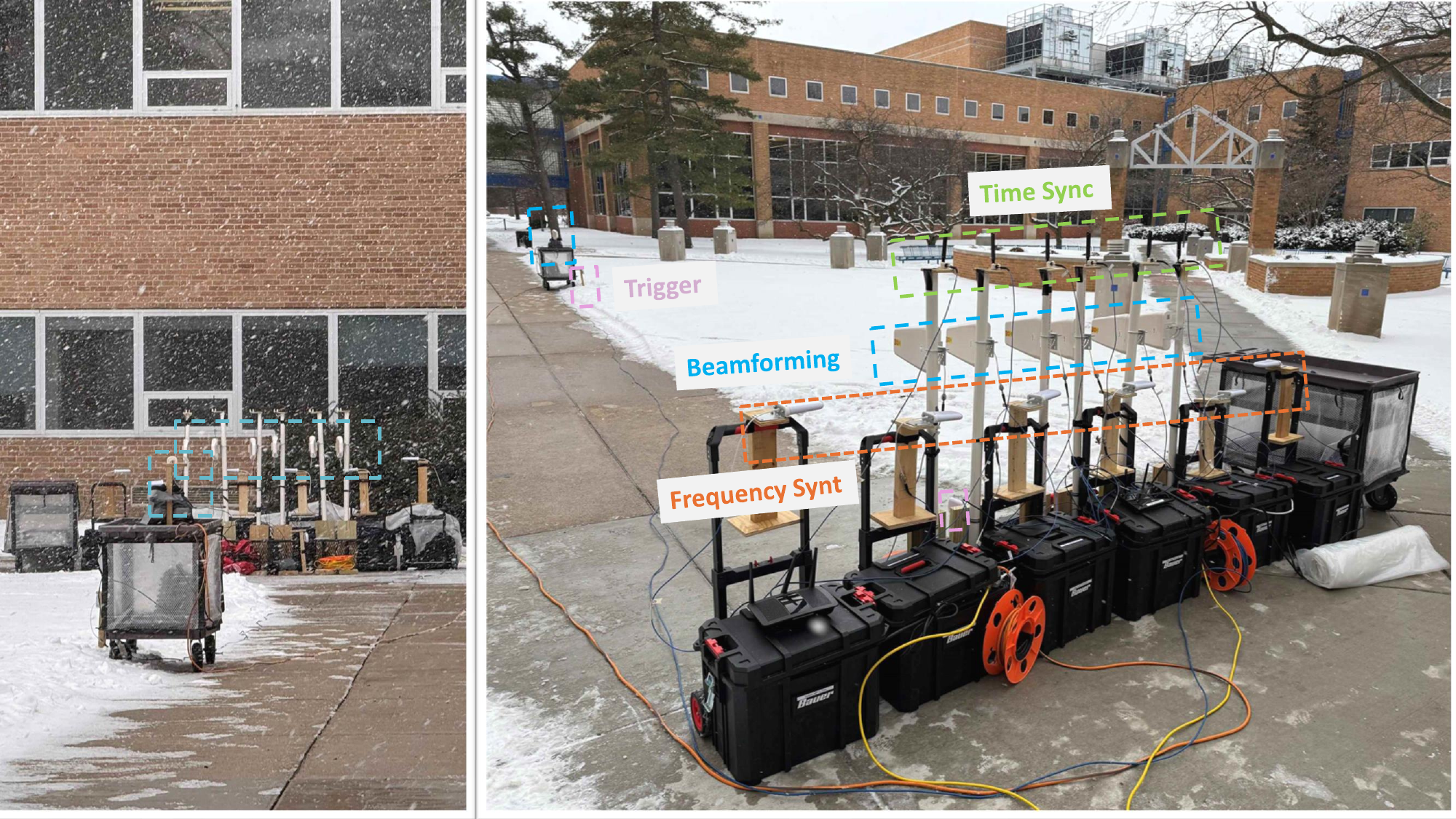}
\caption{Experimental setup for six-node distributed antenna array beamforming. The receiving node consisted of a Keysight oscilloscope connected to an L-Com log-periodic antenna at channel 1 to receive the beamforming pules. A second L-Com directional log-periodic antenna was connected to channel 3 of the oscilloscope for triggering at a frequency of 2 GHz. The receiver was placed 16.3 m from the array.}
\label{setup}
\end{figure*}

\begin{figure*}[t!]
\centering
\includegraphics[width=7in]{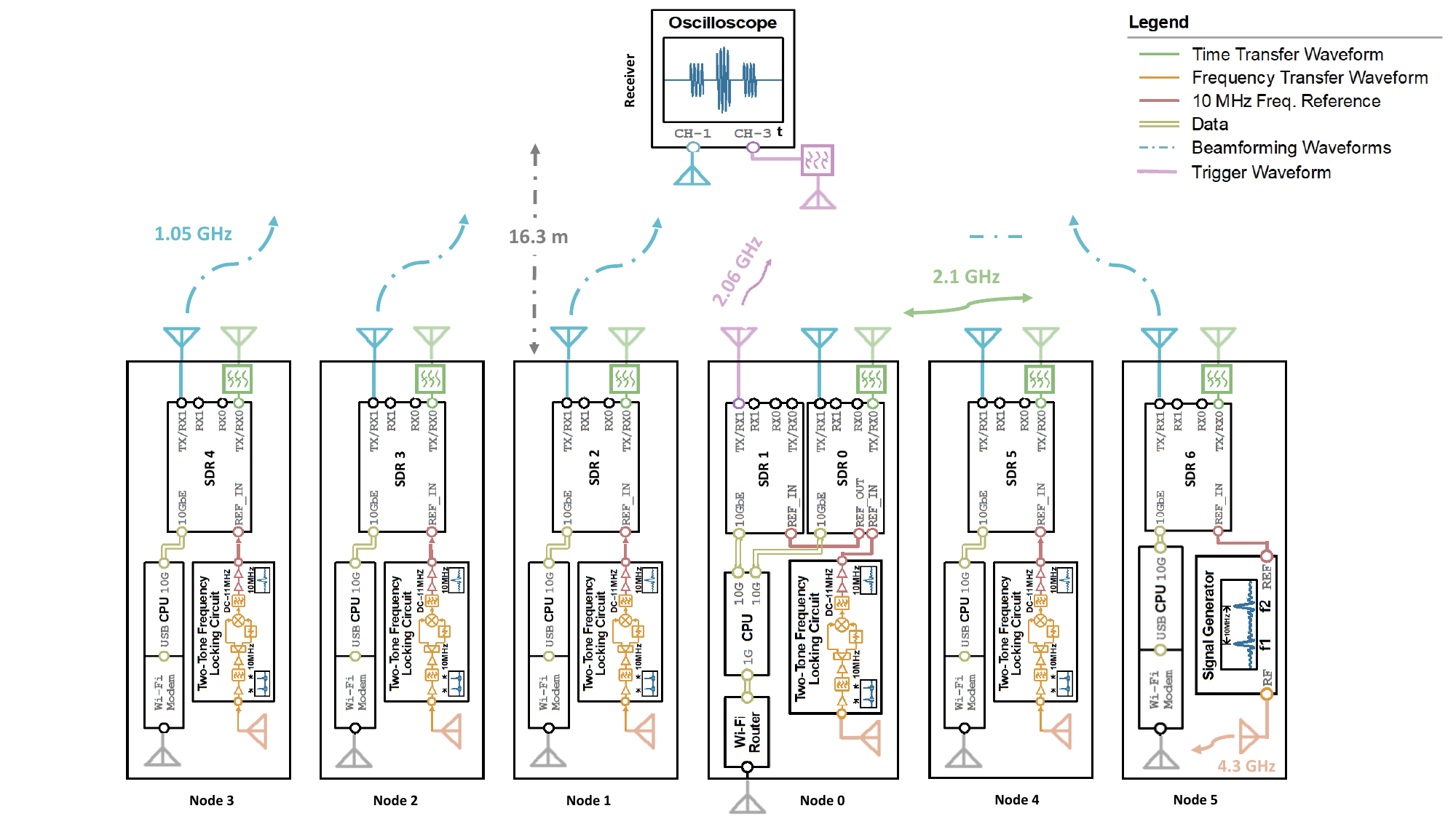}
\caption{Schematic of the coherent beamforming experiment for six-node distributed antenna array. The six SDRs are connected to directional antennas for beamforming and dipole antennas for wireless time transfer at carrier frequency of 2.1 GHz and frequency syntonization at frequency of 4.3 GHz. The system is fully distributed with network initial coarse time alignment.}
\label{sc}
\end{figure*}

The fully wireless six-node antenna array system configuration is detailed in the experimental setup and schematic of Figs.\ \ref{setup} and \ref{sc}, respectively. Each node in the system, housed in a two-wheeled plastic box, consisted of a desktop server as a host computer running Ubuntu 22.04 and equipped with Intel i7-8700s and 16 GB of DDR4 memory. The computer was connected through 10 GbE cable to an Ettus Research X310 Software Defined Radio (SDR) controlled using GNU Radio software and utilized for distributed array coordination on channel 0 and beamforming and beamsteering on channel 1. The system clock on each SDR operated at a rate of 200~MHz and a digital sampling rate of 200~MSa/s, and had two UBX-160 daughterboards supporting instantaneous analog bandwidth of 160~MHz, linked to channels 0 and 1. The Node 0 computer is connected to two SDRs; the second one was used to send a trigger signal to the receiver to record the received signals. The choice of Node 0 for triggering was arbitrary; since the time synchronization process was implemented in a decentralized topology, any node could be used for triggering the receiver. The frequency reference inputs of the SDRs for Nodes 0--4 are connected to the output of the self-mixing frequency locking receiving circuits described in Section \ref{freq_syn} to perform the wireless frequency syntonization. The RF input of each of these circuits was connected to an L-Com HG2458-08LP-NF 8-dBi log-periodic antenna. Node 5 was connected to a Keysight PSG E8267D vector signal generator (VSG) and used as a frequency reference. The VSG output reference frequency was connected directly to the 10 MHz reference on the SDR, and the RF port was connected to an L-Com HG2458-08LP-NF 8-dBi log-periodic antenna to perform the wireless frequency transfer. The VSG was configured to transmit a continuous two-tone signal with 4.3 GHz carrier frequency and 15 dBi amplitude through the log-periodic antenna to be received by the self-mixing circuits in order to provide the 10 MHz frequency references to the other five nodes in the system.


Channel 0 of the SDR was connected to a Taoglas TG.55.8113 wideband monopole antenna covering the 600 to 6000~MHz frequency band and configured to transmit a two-tone LFM waveform with 1 $\mu$s pulse duration, occupied bandwidth of 100\%, and 2.1 GHz carrier frequency, used for the TWTT method employed to achieve decentralized time synchronization as well as inter-node ranging relative to node~0. Selection of waveform parameters for time synchronization was aimed at improving the link SNR values during the beamforming and beamsteering experiments. In previous work, the authors studied the effect of link SNR on the accuracy of time synchronization for a fully connected network~\cite{shandi_conf_2024}. The ambiguity of the two-tone waveform at low link SNR (below 9 dB) results in high errors in the time delay estimation. Therefore, to avoid the ambiguity issue, the two-tone LFM waveform with fully occupied bandwidth was used because of its robustness to noise and multipath and lack of ambiguity at low SNR. To enable network exchange of the timestamps resulting from the TWTT method between the nodes in the array, the system used a distributed computing model. In this model the computation process was performed across the separate host computers and information exchange was accomplished through Wi-Fi (the Node 0 host computer was connected to a Wi-Fi router via 1 GbE, and the other five hosts were connected to Wi-Fi modems).

\begin{figure*}[ht!]
\centering
\includegraphics[width=2\columnwidth]{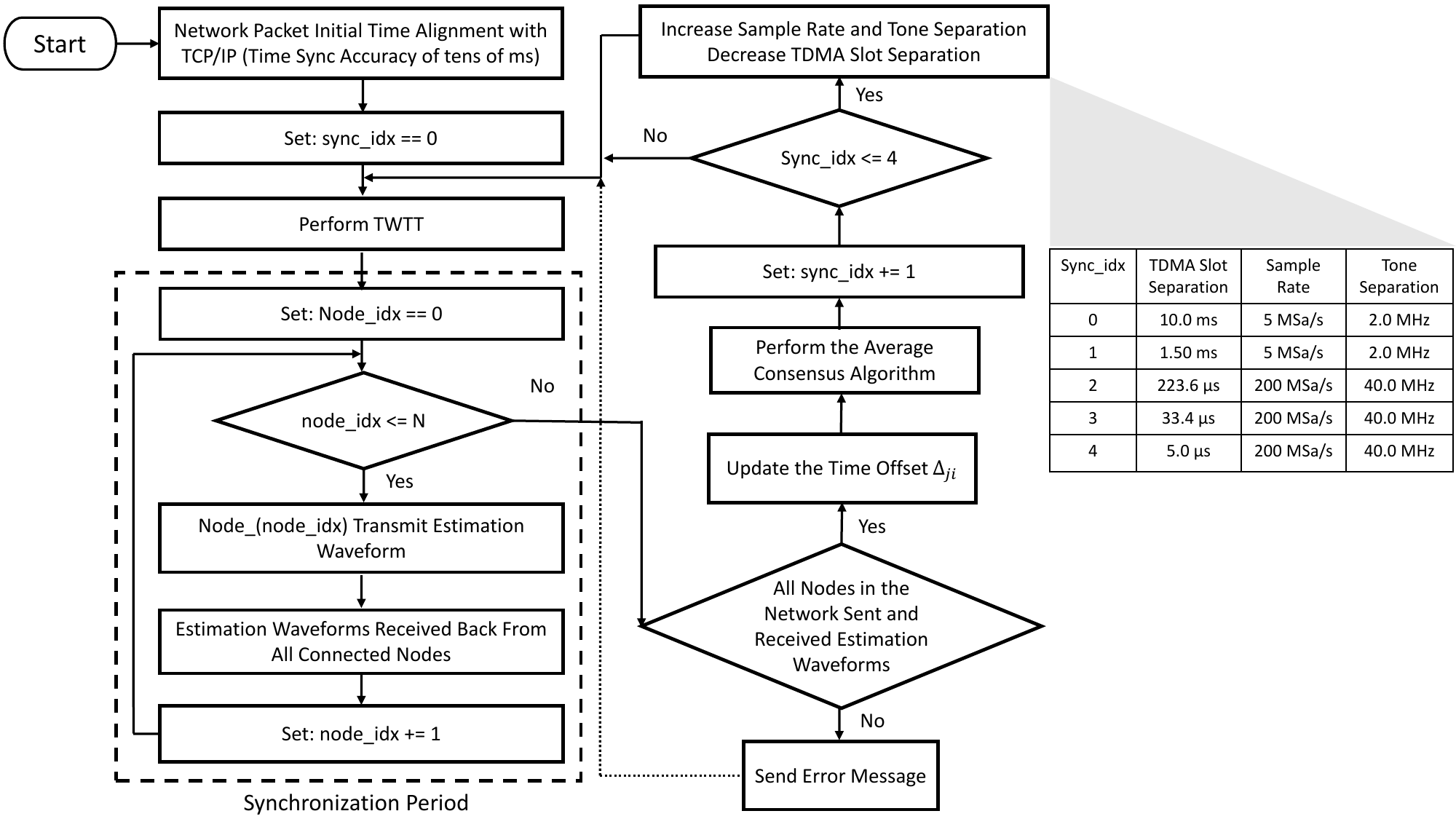}
\caption{Description of the time synchronization stages. The time synchronization process starts with initial network packet time alignment over Wi-Fi, which aligns the clocks of the array to within an accuracy of tens of milliseconds. The initial time alignment is followed by five iterations of iterative refinement, where the sample rate and tone separation increase from 5 MSa/s and 2 MHz to 200 MSa/s and 40 MHz, respectively; in addition, the TDMA slot separation decreases from 10 ms to 5 $\upmu$s to achieve the desired accuracy level needed for the picosecond decentralized time synchronization stage to start (see the table of parameters at lower right). During the last iterative stage, the RF TWTT method is applied and the timestamps are exchanged between the nodes through Wi-Fi. After each iteration, the time offsets between the nodes are updated and the average consensus is applied to compensate for the time bias at the RF front-end. The last stage is accomplished depending on local information exchange between the neighboring nodes without the need for global information.}
\label{fs}
\end{figure*}

The TDMA technique permitted channel sharing among the six nodes by assigning a time slot for each node and allowing only one node to transmit at a time. In order to align the received signals within the defined TDMA window, an initial time alignment was established in the first stage over the network using the TCP/IP technique to achieve time synchronization accuracy on the order of $\sim10$~ms. In the second stage of the time synchronization process, the refinement synchronization method was applied iteratively with sample rate increasing from 5 MSa/s to 200 MSa/s over five iterations. The tone separation increased from 2~MHz to 40 MHz and the TDM window decreased from 10 ms to 5 $\mu$s, reaching the level of alignment allowing for the final decentralized synchronization stage to start achieving a picosecond level of time synchronization accuracy. The flowchart in Fig.\ \ref{fs} details the three-stage time alignment process. 

The last stage of time synchronization was decentralized and depended only on local information sharing between neighboring nodes. In each iteration (synchronization period) the TWTT method was implemented along with the TDMA to ensure channel sharing such that only one node could transmit the time delay estimation waveform at a given time. After all the nodes in the network transmitted, received, and saved the receiving and transmitting timestamps, the host computer exchanged the timestamp information through Wi-Fi. When all the information has arrived at the host of Node 0 (the host chosen arbitrarily to perform the time synchronization final computation step), the average consensus algorithm was applied to compensate for the time bias as described in Section II-C. Finally, channel 1 of the SDR was used to transmit the phase calibration pulses, beamforming pulses, and beamsteering through a 10-dBi L-Com HG72710LP-NF log-periodic antenna.

The receiving node consisted of a four-channel oscilloscope with 20 GSa/s sampling rate. Channel 1 of the oscilloscope was connected to a 10-dBi L-Com HG72710LP-NF log-periodic antenna used to receive the beamforming/beamsteering waveforms, and channel 3 was connected to the same 8-dBi log-periodic antenna used on the transmitter which was used to receive the triggering pulses transmitted by Node 0. A high-$Q$ cavity filter was used to reduce ambient noise and prevent unwanted RF signals from triggering the oscilloscope. The receiving node was located 16.3 m from the six-node array (at broadside in the far field of the array) with nodes positioned in a straight line forming a nonuniform linear array to simplify the calibration process and inter-node ranging. 

\subsection{RESULTS AND DISCUSSION}

\begin{figure}[t!]
\centering
\includegraphics[width=0.9\columnwidth]{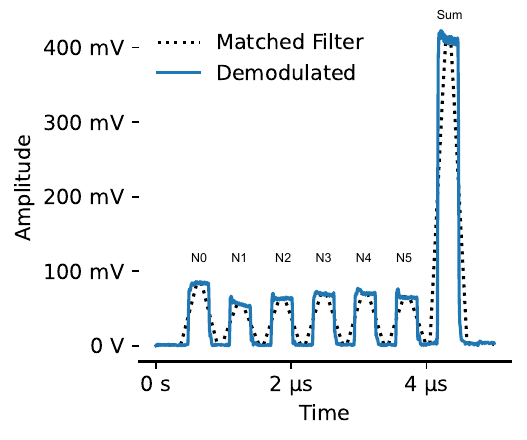}
\caption{One of the 76 samples showing the gain measured by the receiver. The first six pulses in the pulse train represent the individual powers of the six nodes, while the last pulse is the sum of the previous six. The pulse train was received, demodulated, and matched filtered to estimate the peak magnitudes of the pulses. The dashed line shows the matched filter output.}
\label{gc}
\end{figure}

\begin{figure}[t!]
\centering
\includegraphics[width=0.9\columnwidth]{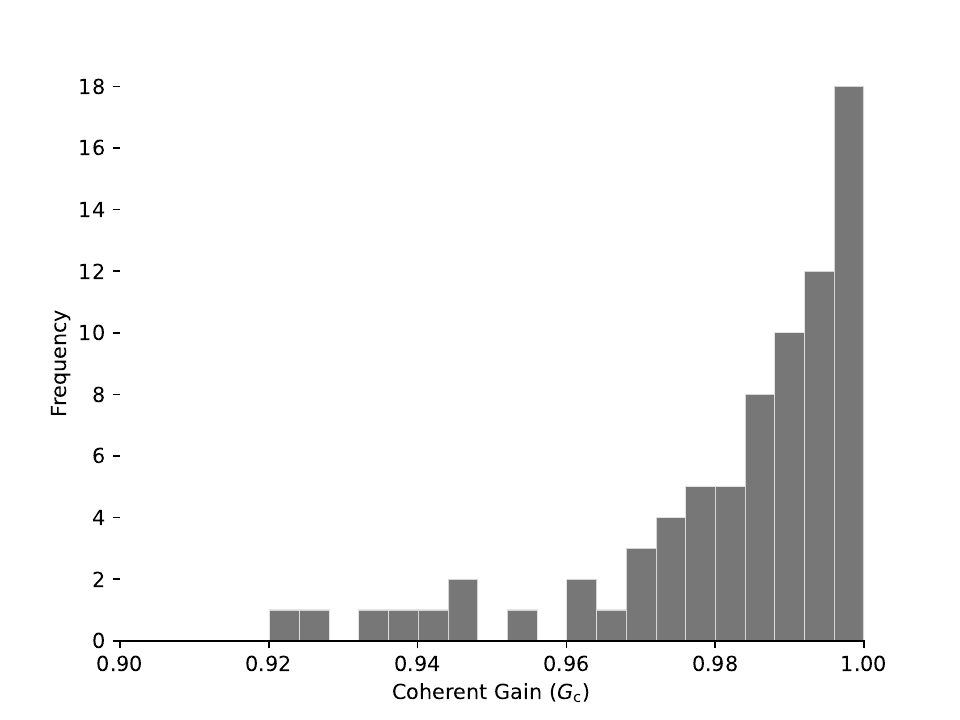}
\caption{Histogram of the measured coherent gain obtained during the beamforming-to-broadside experiment. This histogram was constricted from 76 samples (beamforming waveforms).}
\label{gc_hist}
\end{figure}

The validation process was performed to evaluate the decentralized time synchronization accuracy while focusing on the coherent gain obtained from beamforming and beamsteering. During the experiment beamforming to broadside, 76 samples (pulse trains) were collected to evaluate the gain of the system. Fig.\ \ref{gc} shows one sample of the coherent pulse train received at the destination node. As described in Section III-B and Fig.\ \ref{cw}, the received pulse train consisted of seven pulses with the first six pulses representing the signal powers transmitted by the individual nodes and the last pulse representing the summed power at the destination. The gain was measured by estimating the peak amplitudes of the receiver pulse train. During the estimation process the waveform was downconverted digitally to baseband and then matched filtered to estimate the peaks. Fig.\ \ref{gc_hist} shows the histogram of the coherent gain estimated from the same 76~pulses. The lowest and highest gains were 92\% and $\sim$100\%, respectively, with the mean around 98\%.

\begin{figure}[t!]
\centering
\includegraphics[width=0.9\columnwidth]{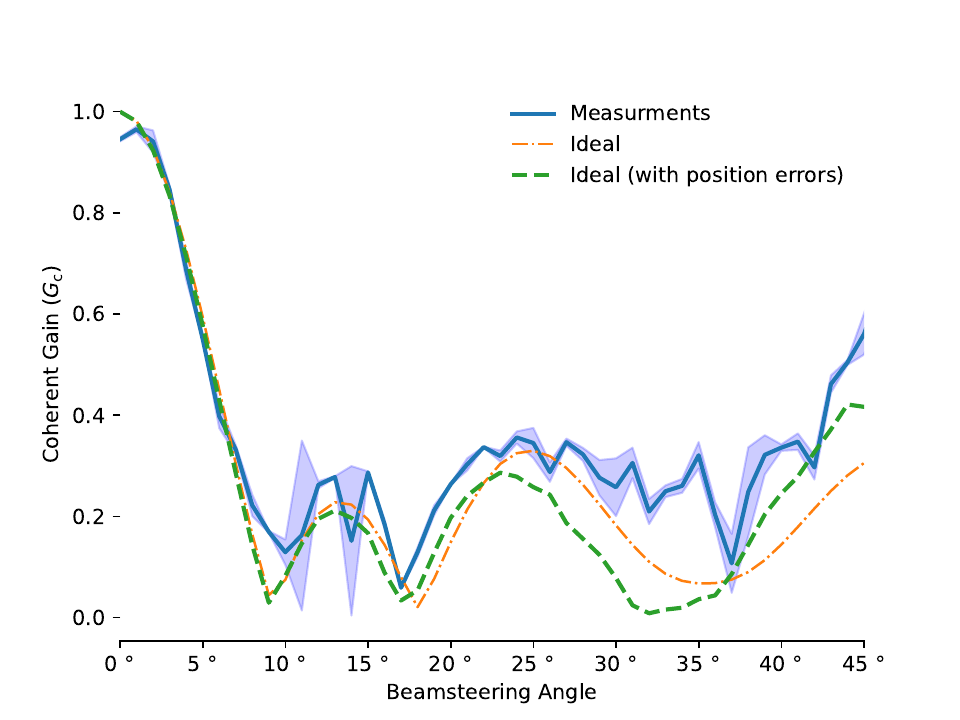}
\caption{Ideal and measured coherent gains during the beamsteering experiment. The ideal gain was simulated twice: first assuming perfect position knowledge and second using the position errors measured by the system during the experiment to obtain a closer approximation for the gain in the presence of position errors. Average gain from two measurements is represented by a solid line, while shaded area shows the max/min values.}
\label{bs}
\end{figure}

Two coherent gain measurements were conducted at each angle while steering the beam from broadside to 45$\degree$, keeping the receiver in static location at broadside. The resulting normalized coherent gain is shown in Fig.~\ref{bs}, where the gain from an ideal array is plotted for comparison in the yellow dash-dot line. Note that as the beam was steered away from broadside, the deviation of the measured system from the ideal gain increased, especially after the beam angle exceeded 30$\degree$. These deviations are believed to be due to multiple factors including position estimation error (as detailed in Table \ref{T1}), frequency jitter, the effect of the cluttered environment, the low signal power, and the formation of grating lobes as the beam was steered to larger angles. Table \ref{T1} details the results from the beamsteering experiment, including the inter-node position estimation errors computed relative to Node 0 with a maximum error of 40~mm. The error increased with the inter-node distance. One source of these errors is the inter-node range calibration bias, caused by measuring the distances between the beamforming antenna phase centers instead of the distances between the TWTT antennas, which were not perfectly aligned with the beamforming antenna phase centers. However, the measured coherent gain pattern was close to the expected ideal array pattern after partial compensation for the position errors, which is shown in the green dashed line in Fig.~\ref{bs}.
\begin{table}[t!]
\begin{center}
\caption{Experimental Results}
\begin{tabular} {ccccc}
    \toprule
    Node & Position Error$^\dagger$&Sync Error$^\dagger$&\multicolumn{2} {c}{Inter-Arrival Delay$^\ddagger$} (ps)\\[2pt]
    \# &Avg (mm)&Avg (ps)& Avg Bias &STD \\ 
   \midrule
    0&ref&  15.76& ref&ref\\
    1&3.12& 19.57 & -8.5&21.6\\
    2&-4.87& 18.09 & 17.8&18.2\\
    3&36.04&35.58 & -6.4&43.8 \\
    4&2.35& 14.09& 13.0&21.4\\
    5&20.04&  16.41& -38.9&23.0\\
    \bottomrule
    \label{T1}
\end{tabular}
\end{center}
\footnotesize{$^\dagger$ Measured at the array.\\$^\ddagger$ Measured at the receiver.}\\
\end{table}

\begin{figure}[t!]
\centering
\includegraphics[width=0.9\columnwidth]{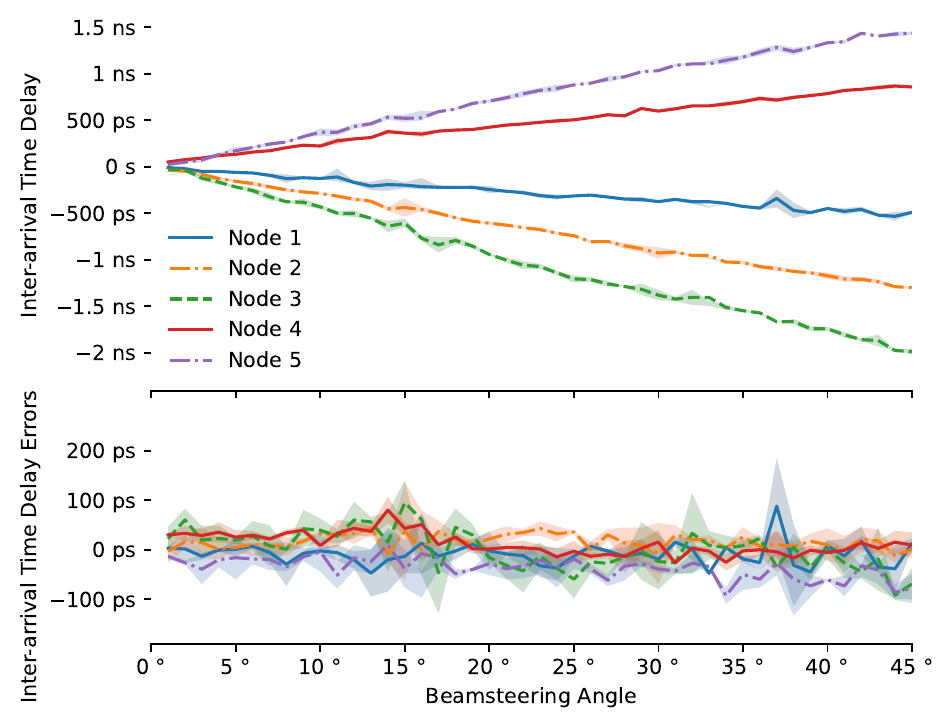}
\caption{Inter-arrival time delays and errors. Top plot shows the estimated inter-arrival time delays, relative to Node 0, of the LFM waveforms transmitted by Nodes 1--5. These delays were estimated by the receiver beamsteering from broadside to 40$^\circ$. Bottom plot shows the deviation of the inter-arrival time delay from the expected delay (the ideal delay resulted from the relative locations of the nodes). Average inter-arrival time delay and the average deviation from two measurements at each angle is represented by solid lines, while shaded areas show the max/min values.}
\label{delay_error}
\end{figure}

\begin{figure}[t!]
\centering
\includegraphics[width=0.9\columnwidth]{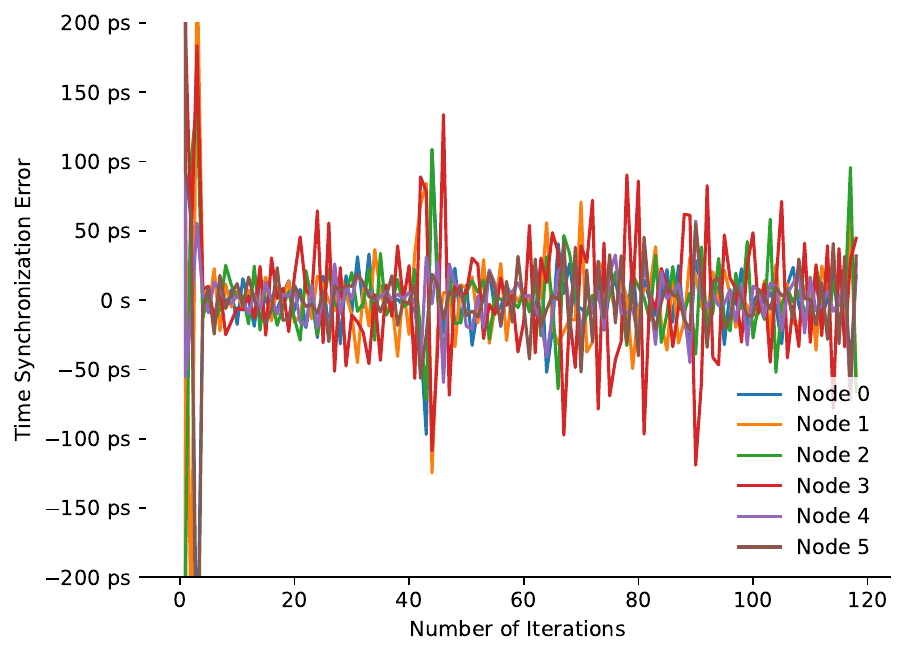}
\caption{Decentralized time synchronization error. These time errors for Nodes 0--5 were estimated during the beamforming experiment and plotted verses the iteration number (synchronization epoch). The average error was computed over 120 iterations for each node, with a maximum average of less than 36 ps.}
\label{time_errors}
\end{figure}

The evaluation of the beamforming inter-node time delay performance is detailed in Fig.\ \ref{delay_error}. The top plot shows the estimated time delays of the LFM waveforms transmitted from Nodes 1--5 and measured at the receiver. These delays were estimated with respect to Node 0 at each beamsteering angle. The bottom plot shows the deviation of the time delays from the ideal delay after compensation for the geometrical delay resulting from the differences in location of the nodes relative to the center of the array. The resulting deviations are caused by the inter-range estimation errors and frequency jitter, which produce distortion of the LFM waveforms. The average bias and the standard deviation of the inter-arrival time delay errors for each pair are detailed in Table \ref{T1}. The maximum average bias was $-38.9$ ps and the maximum standard deviation was $43.8$ ps. Lastly, the decentralized time errors for all six nodes were estimated by computing the error between the predicted and estimated time offsets; these are plotted in Fig.\ \ref{time_errors} over 120~iterations to evaluate the synchronization accuracy. The average error over 120 iterations is computed for each node as shown in Table \ref{T1}. The result shows a maximum error of 150 ps and a maximum average error of less than 36 ps.

\section{CONCLUSIONS}
\label{conclustions}
This work demonstrated distributed beamforming from a fully wireless nonuniform six-node antenna array system to a receiver in the far field (16.3 m). The array coordination was fully wireless, utilizing high accuracy decentralized time synchronization with average error of less than 36 ps. Frequency syntonization was achieved through the implementation of the wireless frequency transfer method. The experimental evaluation included beamsteering in a cluttered environment from broadside to a 45$\degree$ angle. The system achieved an average coherent gain of 98\% of the ideal gain at 1.05 GHz. This showed the system's scalability and demonstrated the applicability of the time synchronization method in the achievement of high time alignment accuracy, thus enabling a high coherent gain for a distributed antenna array with decentralized topology.


\bibliographystyle{IEEEtran}
\bibliography{IEEEabrv, References}

\begin{thebibliography}{10}
\providecommand{\url}[1]{#1}
\csname url@samestyle\endcsname
\providecommand{\newblock}{\relax}
\providecommand{\bibinfo}[2]{#2}
\providecommand{\BIBentrySTDinterwordspacing}{\spaceskip=0pt\relax}
\providecommand{\BIBentryALTinterwordstretchfactor}{4}
\providecommand{\BIBentryALTinterwordspacing}{\spaceskip=\fontdimen2\font plus
\BIBentryALTinterwordstretchfactor\fontdimen3\font minus
  \fontdimen4\font\relax}
\providecommand{\BIBforeignlanguage}[2]{{%
\expandafter\ifx\csname l@#1\endcsname\relax
\typeout{** WARNING: IEEEtran.bst: No hyphenation pattern has been}%
\typeout{** loaded for the language `#1'. Using the pattern for}%
\typeout{** the default language instead.}%
\else
\language=\csname l@#1\endcsname
\fi
#2}}
\providecommand{\BIBdecl}{\relax}
\BIBdecl

\bibitem{4785387}
R.~Mudumbai, D.~R. Brown~Iii, U.~Madhow, and H.~V. Poor, ``Distributed transmit
  beamforming: challenges and recent progress,'' \emph{IEEE Communications
  Magazine}, vol.~47, no.~2, pp. 102--110, 2009.

\bibitem{nanzer2021distributed}
J.~A. Nanzer, S.~R. Mghabghab, S.~M. Ellison, and A.~Schlegel, ``Distributed
  phased arrays: Challenges and recent advances,'' \emph{{IEEE} Trans. Microw.
  Theory Techn.}, vol.~69, no.~11, pp. 4893--4907, 2021.

\bibitem{mi13091481}
\BIBentryALTinterwordspacing
W.~M. Abdulkawi, M.~A. Alqaisei, A.-F.~A. Sheta, and I.~Elshafiey, ``New
  compact antenna array for mimo internet of things applications,''
  \emph{Micromachines}, vol.~13, no.~9, 2022. [Online]. Available:
  \url{https://www.mdpi.com/2072-666X/13/9/1481}
\BIBentrySTDinterwordspacing

\bibitem{10475582}
Z.~Xu, G.~Chen, R.~Fernandez, Y.~Gao, and R.~Tafazolli, ``Enhancement of direct
  {LEO} satellite-to-smartphone communications by distributed beamforming,''
  \emph{IEEE Transactions on Vehicular Technology}, vol.~73, no.~8, pp.
  11\,543--11\,555, 2024.

\bibitem{9798882}
G.~Sun, J.~Li, A.~Wang, Q.~Wu, Z.~Sun, and Y.~Liu, ``Secure and
  energy-efficient {UAV} relay communications exploiting collaborative
  beamforming,'' \emph{IEEE Transactions on Communications}, vol.~70, no.~8,
  pp. 5401--5416, 2022.

\bibitem{9376348}
D.~Tagliaferri, M.~Rizzi, S.~Tebaldini, M.~Nicoli, I.~Russo, C.~Mazzucco, A.~V.
  Monti-Guarnieri, C.~M. Prati, and U.~Spagnolini, ``Cooperative synthetic
  aperture radar in an urban connected car scenario,'' in \emph{2021 1st IEEE
  International Online Symposium on Joint Communications \& Sensing (JC\&S)},
  2021, pp. 1--4.

\bibitem{8378649}
P.~Chatterjee and J.~A. Nanzer, ``Effects of time alignment errors in coherent
  distributed radar,'' in \emph{2018 IEEE Radar Conference (RadarConf18)},
  April 2018, pp. 0727--0731.

\bibitem{4202181}
R.~Mudumbai, G.~Barriac, and U.~Madhow, ``On the feasibility of distributed
  beamforming in wireless networks,'' \emph{IEEE Transactions on Wireless
  Communications}, vol.~6, no.~5, pp. 1754--1763, May 2007.

\bibitem{4542555}
D.~R. {Brown III} and H.~V. {Poor}, ``Time-slotted round-trip carrier
  synchronization for distributed beamforming,'' \emph{IEEE Transactions on
  Signal Processing}, vol.~56, no.~11, pp. 5630--5643, Nov 2008.

\bibitem{4641607}
M.-O. Pun, D.~Richard~Brown, and H.~Vincent~Poor, ``Opportunistic collaborative
  beamforming with one-bit feedback,'' in \emph{2008 IEEE 9th Workshop on
  Signal Processing Advances in Wireless Communications}, 2008, pp. 246--250.

\bibitem{bidgare2012implementation}
P.~Bidigare, M.~Oyarzyn, D.~Raeman, D.~Chang, D.~Cousins, R.~O'Donnell,
  C.~Obranovich, and D.~R. Brown, ``Implementation and demonstration of
  receiver-coordinated distributed transmit beamforming across an ad-hoc radio
  network,'' in \emph{2012 Conference Record of the Forty Sixth Asilomar
  Conference on Signals, Systems and Computers (ASILOMAR)}, 2012, pp. 222--226.

\bibitem{abari2015airshare}
O.~Abari, H.~Rahul, D.~Katabi, and M.~Pant, ``Airshare: Distributed coherent
  transmission made seamless,'' in \emph{2015 IEEE Conference on Computer
  Communications (INFOCOM)}, 2015, pp. 1742--1750.

\bibitem{prager2020wireless}
S.~Prager, M.~S. Haynes, and M.~Moghaddam, ``Wireless subnanosecond {RF}
  synchronization for distributed ultrawideband software-defined radar
  networks,'' \emph{{IEEE} Trans. Microw. Theory Techn.}, vol.~68, no.~11, pp.
  4787--4804, 2020.

\bibitem{alemdar2021rfclock}
K.~Alemdar, D.~Varshney, S.~Mohanti, U.~Muncuk, and K.~Chowdhury, ``{RFClock}:
  timing, phase and frequency synchronization for distributed wireless
  networks,'' in \emph{Proceedings of the 27th Annual International Conference
  on Mobile Computing and Networking}, 2021, pp. 15--27.

\bibitem{merlo2022wireless}
J.~M. Merlo, S.~R. Mghabghab, and J.~A. Nanzer, ``Wireless picosecond time
  synchronization for distributed antenna arrays,'' \emph{IEEE Transactions on
  Microwave Theory and Techniques}, vol.~71, no.~4, pp. 1720--1731, 2022.

\bibitem{wang}
\BIBentryALTinterwordspacing
D.~Lucarelli and I.-J. Wang, ``Decentralized synchronization protocols with
  nearest neighbor communication,'' in \emph{Proceedings of the 2nd
  International Conference on Embedded Networked Sensor Systems}, ser. SenSys
  '04.\hskip 1em plus 0.5em minus 0.4em\relax New York, NY, USA: Association
  for Computing Machinery, 2004, p. 62–68. [Online]. Available:
  \url{https://doi.org/10.1145/1031495.1031503}
\BIBentrySTDinterwordspacing

\bibitem{ouassal2021decentralized}
H.~Ouassal, M.~Yan, and J.~A. Nanzer, ``Decentralized frequency alignment for
  collaborative beamforming in distributed phased arrays,'' \emph{{IEEE} Trans.
  Wireless Commun.}, vol.~20, no.~10, pp. 6269--6281, 2021.

\bibitem{140489}
Y.~Akaiwa, H.~Andoh, and T.~Kohama, ``Autonomous decentralized
  inter-base-station synchronization for tdma microcellular systems,'' in
  \emph{[1991 Proceedings] 41st IEEE Vehicular Technology Conference}, 1991,
  pp. 257--262.

\bibitem{10600440}
R.~H. Kenney and J.~W. McDaniel, ``All-digital carrier frequency
  synchronization for distributed radar sensor networks,'' in \emph{2024
  IEEE/MTT-S International Microwave Symposium - IMS 2024}, 2024, pp. 493--496.

\bibitem{10443654}
R.~H. Kenney, J.~G. Metcalf, and J.~W. McDaniel, ``Wireless distributed
  frequency and phase synchronization for mobile platforms in cooperative
  digital radar networks,'' \emph{IEEE Transactions on Radar Systems}, vol.~2,
  pp. 268--287, 2024.

\bibitem{merlo2024past}
J.~M. Merlo, N.~Shandi, M.~Dula, A.~Bhattacharyya, and J.~A. Nanzer, ``Fully
  wireless collaborative beamforming using a three-element coherent distributed
  phased array,'' in \emph{2024 IEEE International Symposium on Phased Array
  Systems and Technology (ARRAY)}, 2024, pp. 1--8.

\bibitem{shandi2024}
N.~Shandi, J.~M. Merlo, and J.~A. Nanzer, ``Decentralized picosecond
  synchronization for distributed wireless systems,'' \emph{IEEE Transactions
  on Communications}, pp. 1--1, 2024.

\bibitem{pozar2005microwave}
D.~Pozar, \emph{Microwave Engineering}, 3rd~ed.\hskip 1em plus 0.5em minus
  0.4em\relax John Wiley \& Sons, Inc., 2005.

\bibitem{levine2008review}
J.~Levine, ``A review of time and frequency transfer methods,''
  \emph{Metrologia}, vol.~45, no.~6, p. S162, 2008.

\bibitem{boyd}
\BIBentryALTinterwordspacing
S.~Boyd, P.~Diaconis, and L.~Xiao, ``Fastest mixing markov chain on a graph,''
  \emph{SIAM Review}, vol.~46, no.~4, pp. 667--689, 2004. [Online]. Available:
  \url{https://doi.org/10.1137/S0036144503423264}
\BIBentrySTDinterwordspacing

\bibitem{8649676}
H.~Abdzadeh-Ziabari, W.-P. Zhu, and M.~N.~S. Swamy, ``Timing and frequency
  synchronization and doubly selective channel estimation for {OFDMA} uplink,''
  \emph{IEEE Transactions on Circuits and Systems II: Express Briefs}, vol.~67,
  no.~1, pp. 62--66, 2020.

\bibitem{8058715}
M.~Pontón and A.~Suárez, ``Stability analysis of wireless coupled-oscillator
  circuits,'' in \emph{2017 IEEE MTT-S International Microwave Symposium
  (IMS)}, 2017, pp. 83--86.

\bibitem{6848566}
X.~Yang, X.~Lu, and A.~Babakhani, ``Picosecond wireless synchronization using
  an optically locked voltage controlled oscillator {(OL-VCO)},'' in \emph{2014
  IEEE MTT-S International Microwave Symposium (IMS2014)}, 2014, pp. 1--4.

\bibitem{4319216}
K.-Y. Tu and C.-S. Liao, ``Application of {ANFIS} for frequency syntonization
  using {GPS} carrier-phase measurements,'' in \emph{2007 IEEE International
  Frequency Control Symposium Joint with the 21st European Frequency and Time
  Forum}, 2007, pp. 933--936.

\bibitem{mghabghab2021open-loop}
S.~R. Mghabghab and J.~A. Nanzer, ``Open-loop distributed beamforming using
  wireless frequency synchronization,'' \emph{{IEEE} Trans. Microw. Theory
  Techn.}, vol.~69, no.~1, pp. 896--905, 2021.

\bibitem{shandi_conf_2024}
N.~Shandi, J.~M. Merlo, and J.~A. Nanzer, ``High accuracy decentralized time
  synchronization using {SNR} based weighting,'' in \emph{2024 IEEE
  International Symposium on Antennas and Propagation and INC/USNC‐URSI Radio
  Science Meeting (AP-S/INC-USNC-URSI)}, 2024, pp. 1873--1874.

\end{thebibliography}

\newpage

\begin{IEEEbiography}[{\includegraphics[width=1in,height=1.25in,clip,keepaspectratio]{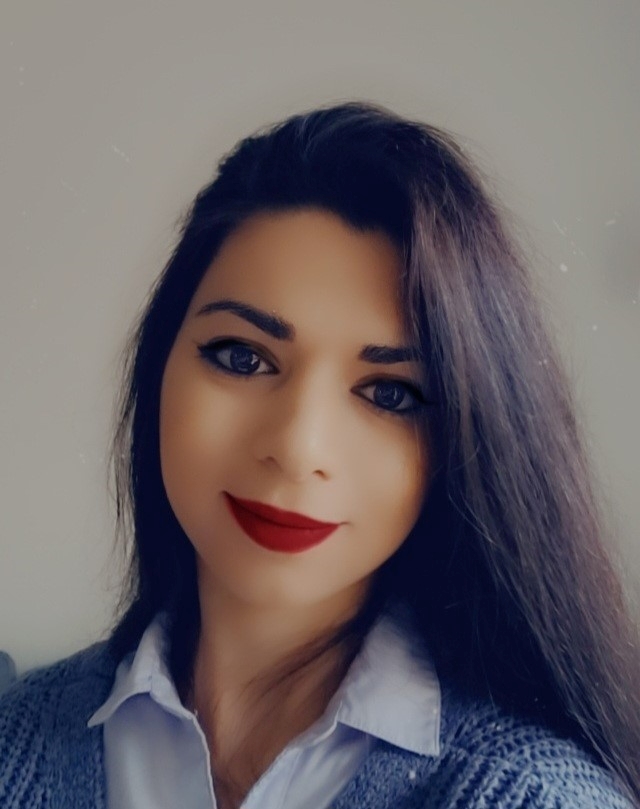}}]{Naim Shandi}{\space}(Graduate Student Member, IEEE) received the B.S. degree in electrical engineering, and the M.S. degree in electrical and computer engineering, from Lawrence Technological University, Southfield, MI, USA, in 2020 and 2022, respectively.

From 2019 to 2022, she was a Technical Test Engineer at Gentherm Incorporated, Farmington Hills, MI, USA, where she planned and carried out electromagnetic compatibility testing procedures. In 2022, she joined the Department of Electrical and Computer Engineering, Michigan State University, East Lansing, MI, USA, where she is presently a graduate research assistant. Her current research interests include distributed arrays, millimeter-wave systems, radar and signal processing, microwave wireless systems, and electromagnetic material characterization.
\end{IEEEbiography}

\begin{IEEEbiography}[{\includegraphics[width=1in,height=1.25in,clip,keepaspectratio]{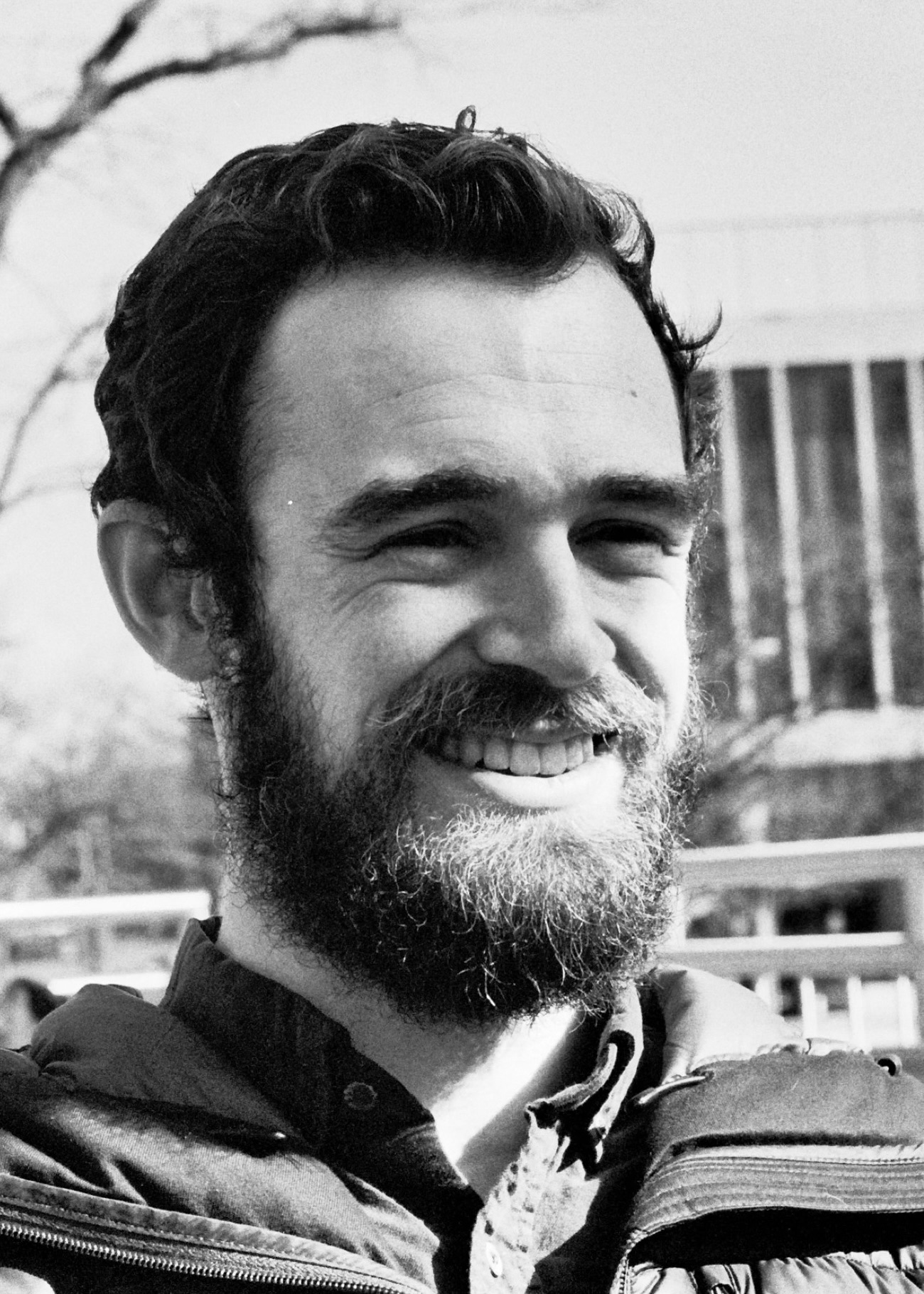}}]{Jason M. Merlo}{\space}(Graduate Student Member, IEEE) received the B.S. degree in computer engineering from Michigan State University, East Lansing, MI, USA in 2018, where he is currently a Ph.D. candidate in electrical engineering.

From 2017-2021 he was project manager and electrical systems team lead of the Michigan State University AutoDrive Challenge team. His current research interests include software-defined radios, distributed radar and wireless system synchronization, interferometric arrays, synthetic aperture radar, joint radar-communications, and automotive radar applications.

Jason was a Finalist in the Student Paper Competition at the IEEE International Microwave symposium in 2024 and won First Place in 2023. He was a Finalist and Honorable Mention in the Student Paper Competition at the IEEE International Symposium on Antennas and Propagation in 2024 and 2023, respectively. He was the recipient of a 2023 IEEE MTT-S Graduate Fellowship and a 2023 URSI Young Scientist Award. He won Second Place in the Student Design Competition at the 2020 IEEE International Symposium on Antennas and Propagation and was a finalist in the Student Safety Technology Design Competition at the 2017 International Technical Conference on Enhanced Safety of Vehicles.
\end{IEEEbiography}

\begin{IEEEbiography}[{\includegraphics[width=1in,height=1.25in,clip,keepaspectratio]{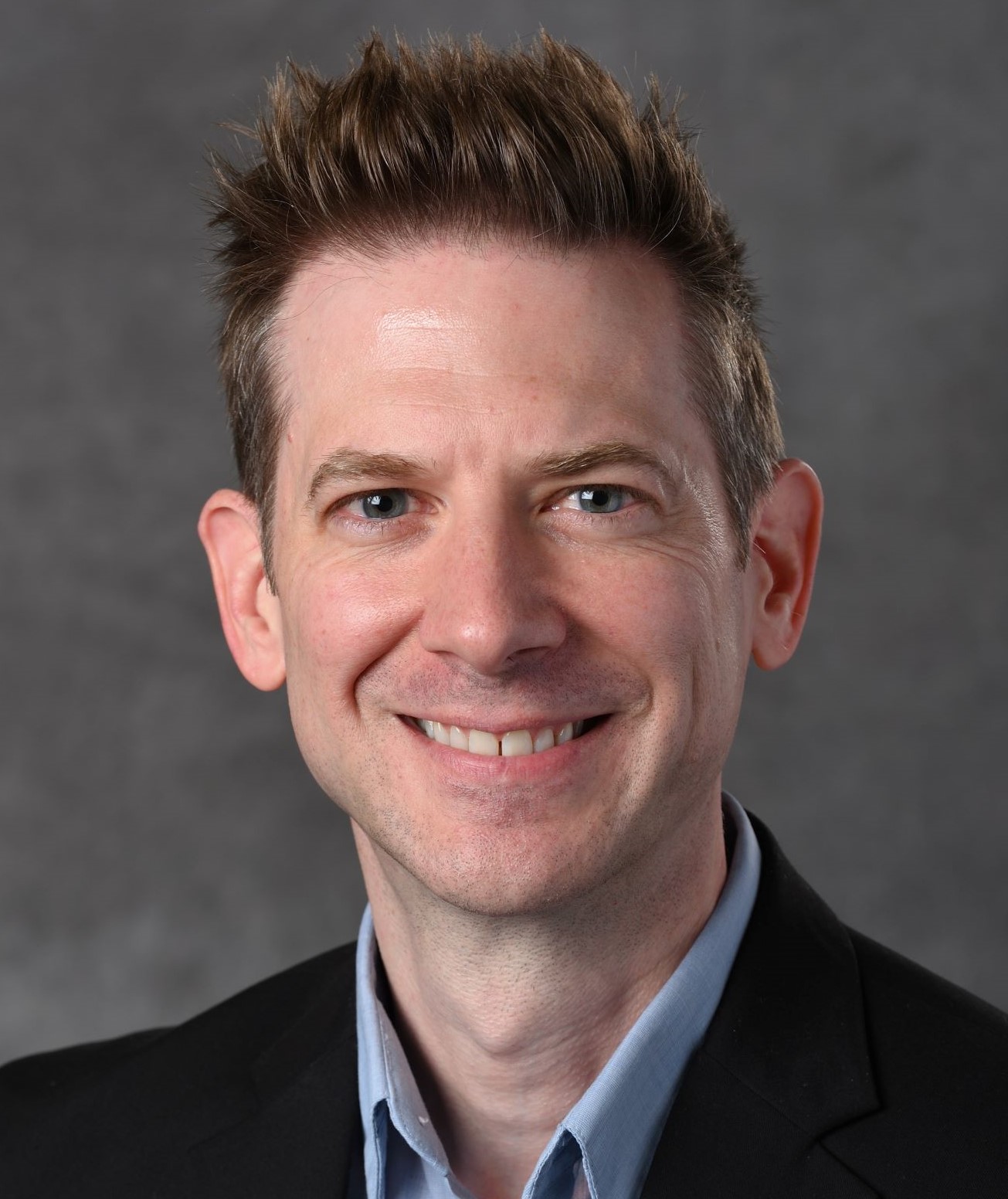}}]{Jeffrey A. Nanzer}{\space} (Senior Member, IEEE) received the B.S. degrees in electrical engineering and in computer engineering from Michigan State University, East Lansing, MI, USA, in 2003, and the M.S. and Ph.D. degrees in electrical engineering from The University of Texas at Austin, Austin, TX, USA, in 2005 and 2008, respectively.

From 2008 to 2009 he was with the University of Texas Applied Research Laboratories in Austin, Texas as a Post-Doctoral Fellow designing electrically small HF antennas and communications systems. From 2009 to 2016 he was with the Johns Hopkins University Applied Physics Laboratory where he created and led the Advanced Microwave and Millimeter-Wave Technology Section. In 2016 he joined the Department of Electrical and Computer Engineering at Michigan State University where he held the Dennis P. Nyquist Assistant Professorship from 2016 through 2021. He is currently an Associate Professor. He directs the Electromagnetics Laboratory, which consists of the Antenna Laboratory, the Radar Laboratory, and the Wireless Laboratory. He has published more than 250 refereed journal and conference papers, two book chapters, and the book  Microwave and Millimeter-Wave Remote Sensing for Security Applications (Artech House, 2012). His research interests are in the areas of distributed phased arrays, dynamic antenna arrays, millimeter-wave imaging, remote sensing, millimeter-wave photonics, and electromagnetics.

Dr. Nanzer is a Distinguished Microwave Lecturer for the IEEE Microwave Theory and Techniques Society (Tatsuo Itoh Class of 2022-2024). He was a Guest Editor of the Special Issue on Special Issue on Radar and Microwave Sensor Systems in the IEEE Microwave and Wireless Components Letters in 2022. He is a member of the IEEE Antennas and Propagation Society Education Committee and the USNC/URSI Commission B, was a founding member and the First Treasurer of the IEEE APS/MTT-S Central Texas Chapter, served as the Vice Chair for the IEEE Antenna Standards
Committee from 2013 to 2015, and served as the Chair of the Microwave Systems
Technical Committee (MTT-16), IEEE Microwave Theory and Techniques
Society from 2016 to 2018.
He was a recipient of the Withrow Junior Distinguished Scholar Award in 2024, the Google Research Scholar Award in 2022 and 2023, the IEEE MTT-S Outstanding
Young Engineer Award in 2019, the DARPA Director’s Fellowship in 2019, the National Science Foundation (NSF) CAREER Award in 2018, the DARPA Young
Faculty Award in 2017, and the JHU/APL Outstanding Professional Book
Award in 2012.

\end{IEEEbiography}
\end{document}